\documentclass{eptcs-modified}
\usepackage{breakurl}             

\usepackage{amsmath}
\usepackage{amssymb}
\usepackage{url}
\usepackage{amsthm}
\usepackage{graphicx}
\usepackage{amscd}

\usepackage{tikz} 
\usepackage[boxed,linesnumbered,longend]{algorithm2e} 
\usetikzlibrary{positioning}
\usetikzlibrary{shapes.geometric}

\usetikzlibrary{automata}
\usepackage{subcaption}

\newcommand{\name}[1]{\textsc{#1}}
\newcommand{\hide}[1]{}

\theoremstyle{definition}
\newtheorem{theorem}{Theorem}
\newtheorem{definition}[theorem]{Definition}

\newtheorem{corollary}[theorem]{Corollary}
\newtheorem{proposition}[theorem]{Proposition}
\newtheorem{lemma}[theorem]{Lemma}
\newtheorem{observation}[theorem]{Observation}

\newtheorem{question}[theorem]{Open Question}

\newtheorem{example}[theorem]{Example}
\newtheorem*{lemma*}{Lemma}
\newtheorem*{observation*}{Observation}
\newcommand{\uint}{{[0,1]}}

\newcommand{\coloring}{\Gamma}

\title{Extending Finite Memory Determinacy to Multiplayer Games}
\author{St\'ephane Le Roux
\institute{D\'epartement d'informatique\\ Universit\'e libre de
Bruxelles, Belgique}
\email{Stephane.Le.Roux@ulb.ac.be}
\and
Arno Pauly
\institute{ D\'epartement d'Informatique\\ Universit\'e Libre de Bruxelles, Belgium
\email{Arno.Pauly@cl.cam.ac.uk}}
}
\def\titlerunning{Extending finite memory determinacy}
\def\authorrunning{S.~Le Roux \& A.~Pauly}
\begin{document}
\maketitle

\begin{abstract}
We show that under some general conditions the finite memory determinacy of a class of two-player win/lose games played on finite graphs implies the existence of a Nash equilibrium built from finite memory strategies for the corresponding class of multi-player multi-outcome games. This generalizes a previous result by  Brihaye, De Pril and Schewe. We provide a number of example that separate the various criteria we explore.

Our proofs are generally constructive, that is, provide upper bounds for the memory required, as well as algorithms to compute the relevant Nash equilibria.
\end{abstract}

\section{Introduction}
The usual model employed for synthesis are sequential two-player win/lose games played on finite graphs. The vertices of the graph correspond to states of a system, and the two players jointly generate an infinite path through the graph (the \emph{run}). One player, the protagonist, models the aspects of the system under the control of the designer. In particular, the protagonist will win the game iff the run satisfies the intended specification. The other player is assumed to be fully antagonistic, thus wins iff the protagonist loses. One then would like to find winning strategies of the protagonist, that is, a strategy for her to play the game in such a way that she will win regardless of the antagonist's moves. Particularly desirable winning strategies are those which can be executed by a finite automaton.

Classes of games are distinguished by the way the winning conditions (or more generally, preferences of the players) are specified. Typical examples include:
\begin{itemize}
\item Muller conditions, where only the set of vertices visited infinitely many times matters;
\item Parity conditions, where each vertex has a priority, and the winner is decided by the parity of the least priority visited infinitely many times;
\item Energy conditions, where each vertex has an energy delta (positive or negative), and the protagonist loses if the cumulative energy values ever drop below 0;
\item Discounted payoff conditions, where each vertex has a payoff value, and the outcome is determined by the discounted sum of all payoffs visited with some discount factor $0 < \lambda < 1$;
\item Combinations of these, such as energy parity games, where the protagonist has to simultaneously ensure that the least parity visited infinitely many times is odd and that the cumulative energy value is never negative.
\end{itemize}

Our goal is to dispose of two restrictions of this setting: First, we would like to consider any number of players; and second allow them to have far more complicated preferences than just preferring winning over losing. The former generalization is crucial in a distributed setting (also e.g.~\cite{depril2,bulling}): If different designers control different parts of the system, they may have different specifications they would like to enforce, which may be partially but not entirely overlapping. The latter seems desirable in a broad range of contexts. Indeed, rarely is the intention for the behaviour of a system formulated entirely in black and white: We prefer a programm just crashing to accidently erasing our hard-drive; we prefer a programm to complete its task in 1 minute to it taking 5 minutes, etc. We point to \cite{kupferman} for a recent survey on such notions of quality in synthesis.

Rather than achieving this goal by revisiting each individual type of game and proving the desired results directly (e.g.~by generalizing the original proofs of the existence of winning strategies), we shall provide two transfer theorems: In both Theorem \ref{thm:fmNE-many-games} and Theorem \ref{thm:fmNE-one-game-bit}, we will show that (under some conditions), if the two-player win/lose version of a game is finite-memory determined, the corresponding multi-player multi-outcome games all have finite-memory Nash equilibria. The difference is that Theorem \ref{thm:fmNE-many-games} refers to all games played on finite graphs using certain preferences, whereas Theorem \ref{thm:fmNE-one-game-bit} refers to one fixed graph only.

Theorem \ref{thm:fmNE-one-game-bit} is more general than a similar one obtained by \name{Brihaye}, \name{De Pril} and \name{Schewe} \cite{depril2},\cite[Theorem 4.4.14]{depril}. A particular class of games covered by our result but not the previous one are (a variant of) energy parity games as introduced by \name{Chatterjee} and \name{Doyen} \cite{chatterjee4}. The high-level proof idea follows earlier work by the authors on equilibria in infinite sequential games, using Borel determinacy as a blackbox \cite{paulyleroux2}\footnote{Precursor ideas are also present in \cite{leroux3} and \cite{mertens} (the specific result in the latter was joint work with Neymann).} -- unlike the constructions there (cf.~\cite{paulyleroux3-cie}), the present ones however are constructive and thus give rise to algorithms computing the equilibria in the multi-player multi-outcome games given suitable winning strategies in the two-player win/lose versions.

The general theme of transferring determinacy results from antagonistic two-player games to the existence of Nash equilibria in multiplayer games is already present in \cite{thuijsman} by \name{Thuijsman} and \name{Raghavan}, as well as \cite{ummels2} by \name{Gr\"adel} and \name{Ummels}.

Echoing \name{De Pril} in \cite{depril}, we would like to stress that our conditions apply to the preferences of each player individually. For example, some players could pursue energy parity conditions, whereas others have preferences based on Muller conditions: Our results apply just as they would do if all players had preferences of the same type.

This article extends and supersedes the earlier \cite{paulyleroux4-sr} which appeared in the proceedings of Strategic Reasoning 2016.

{\bf Structure of the paper:} After introducing notation and the basic concepts in Section \ref{sec:background}, we state our two main theorems in Section \ref{sec:mainresults}. The proofs of our main theorems are given in the form of several lemmata in Section \ref{sec:proofs}. The lemmata prove slightly more than required for the theorems, and might be of independent interest for some readers. In Section \ref{section:discussion} we discuss how our results improve upon prior work, and explore several notions prominent in our main theorems in some more detail. Finally, in Section \ref{sec:applications} we consider as applications two classes of games covered by our main theorems but not by previous work.

\section{Background}
\label{sec:background}

\paragraph{Win/lose two-player games:}

A win/lose two-player game played on a finite graph is specified by a directed graph $(V,E)$ where every vertex has an outgoing edge, a starting vertex $v_0\in V$, two sets  $V_1 \subseteq V$ and $V_2 := V \setminus V_1$, a function $\coloring: V \to C$ coloring the vertices, and a \emph{winning condition} $W\subseteq C^\omega$. Starting from  $v_0$, the players move a token along the graph, $\omega$ times, with player $a\in\{1,2\}$ picking and following an outgoing edge whenever the current vertex lies in $V_a$. Player $1$ wins iff the infinite sequence of the colors seen (at the visited vertices) is in $W$.

\paragraph{Winning strategies:}

For $a\in\{1,2\}$ let $\mathcal{H}_a$ be the set of finite paths in $(V,E)$ starting at $v_0$ and ending in $V_a$. Let $\mathcal{H} := \mathcal{H}_1 \cup \mathcal{H}_2$ be the possible \emph{finite histories} of the game, and let $[\mathcal{H}]$ be the infinite ones. For clarity we may write $[\mathcal{H}_g]$ instead of $[\mathcal{H}]$ for a game $g$. A \emph{strategy} of player $a\in\{1,2\}$ is a function of type $\mathcal{H}_a \to V$ such that $(v,s(hv)) \in E$ for all $hv\in \mathcal{H}_a$. A pair of strategies $(s_1,s_2)$ for the two players induces a run $\rho \in V^\omega$: Let $s := s_1 \cup s_2$ and set $\rho (0) := v_0$ and $\rho (n+1) := s(\rho(0)\rho(1)\ldots \rho(n))$. For all strategies $s_a$ of player $a$ let $\mathcal{H}(s_a)$ be the finite histories in $\mathcal{H}$ that are compatible with $s_a$, and let $[\mathcal{H}(s_a)]$ be the infinite ones. Every sequence $v_0v_1v_2\dots$ of vertices naturally induces a color trace $\coloring(v_0v_1v_2\dots) := \coloring(v_0)\coloring(v_1)\coloring(v_2)\dots$. A strategy $s_a$ is said to be winning if $\coloring[[\mathcal{H}(s_a)]] \subseteq W$, \textit{i.e.} $a$ wins regardless of her opponent's moves. A game where either of the players has a winning strategy is called \emph{determined}.

\paragraph{Finite-memory strategies:}

A \emph{strategic update} for player $a$ using $m$ bits of memory is a function $\sigma: V \times \{0,1\}^m \to V\times \{0,1\}^m$ that describes the two simultaneous updates of player $a$ upon arrival at a vertex $v$ if its memory content was $M$ just before arrival: $(v,M)\mapsto \pi_2\circ\sigma(v,M)$ describes the memory update and $(v,M)\mapsto \pi_1\circ\sigma(v,M)$ the choice for the next vertex. This choice will be ultimately relevant only if $v\in V_a$, in which case we require that $(v,\pi_1\circ\sigma(v,M)) \in E$.

A strategic update together with an initial memory content $M_\epsilon\in \{0,1\}^m$ is called a \emph{strategic implementation}. The memory content after some history is defined by induction: $M_{\sigma}(M_\epsilon,\epsilon):=M_\epsilon$ and $M_{\sigma}(M_\epsilon,hv) := \pi_2 \circ \sigma(v,M_{\sigma}(M_\epsilon,h))$ for all $hv\in \mathcal{H}$. A strategic update $\sigma$ together with initial memory content $M_\epsilon$ induce a \emph{finite-memory strategy} $s_a$ defined by $s_a(hv):= \pi_1\circ\sigma(v,M_{\sigma}(M_\epsilon,h))$ for all $hv\in \mathcal{H}_a$. In a slight abuse we may call strategic implementation finite-memory strategies. If not stated otherwise, we will assume the initial memory to be $0^m$.

\paragraph{Multi-outcome multi-player games and Nash equilibria:}

A (general) game played on a finite graph is specified by a directed graph $(V,E)$, a set of agents $A$, a cover $\{V_a\}_{a \in A}$ of $V$ via pairwise disjoint sets, the starting vertex $v_0$, a function $\coloring: V \to C$ coloring the vertices, and for each player $a$ a preference relation\footnote{Note that we do not understand preferences to automatically be total or satisfy other specific properties. From Definition \ref{def:strictweakorder} onwards we will restrict our attention to strict weak orders though.} $\mathalpha{\prec}_a \subseteq \coloring[[\mathcal{H}]] \times  \coloring[[\mathcal{H}]]$ (or more generously $\mathalpha{\prec}_a \subseteq C^\omega \times C^\omega$). We overload the notation by also writing $\rho \prec_a \rho'$ if $\coloring(\rho) \prec_a \coloring(\rho')$ for all $\rho,\rho' \in [\mathcal{H}]$, \textit{i.e.} players compare runs by comparing their color traces. The two-player games with inverse preferences ($\mathalpha{\prec}_2 = \mathalpha{\prec}_1^{-1}$) are called antagonistic games, and they generalize win/lose two-player games.

The notions of strategies and induced runs generalize in the obvious way. In particular, instead of a pair of strategies (one per player), we consider families $(s_a)_{a \in A}$, which are called strategy profiles. The concept of a winning strategy no longer applies though. Instead, we use the more general notion of a Nash equilibrium: A family of strategies $(s_a)_{a \in A}$ is a Nash equilibrium, if there is no player $a_0 \in A$ and strategy $s'_{a_0}$ such that $a_0$ would prefer the run induced by $(s_a)_{a \in A \setminus \{a_0\}} \cup (s'_{a})_{a \in \{a_0\}}$ to the run induced by $(s_a)_{a \in A}$. Intuitively, no player can gain by unilaterally deviating from a Nash equilibrium. Note that the Nash equilibria in two-player win/lose games are precisely those pairs of strategy where one strategy is a winning strategy.

\paragraph{Threshold games and future games:}

Our results, including transfer from the two-player win/lose case to the general case, rely on the idea that each general game induces a collection of two-player win/lose games, namely the threshold games of the future games, as below.

\begin{definition}[Future game and one-vs-all threshold game] \hfill

Let $g = \langle (V,E), v_0, A, \{V_a\}_{a \in A}, (\prec_a)_{a \in A}\rangle$ be a game played on a finite graph.
\begin{itemize}
\item For $a_0\in A$ and $\rho \in [\mathcal{H}]$, the one-vs-all threshold game $g_{a_0,\rho}$ for $a_0$ and $\rho$ is the win-lose two-player game played on $(V,E)$, starting at $v_0$, with vertex subsets  $V_{a_0}$ and $\bigcup_{a \in A \setminus \{a_0\}} V_a$, and with winning set $\{\rho'\in [\mathcal{H}]\,\mid\,\rho \prec_{a_0} \rho'\}$ for Player $1$.

\item Let $v\in V$. For paths $hv$ and $vh'$ in $(V,E)$ let $hv\hat{\,}vh' := hvh'$.

\item For all $h\in \mathcal{H}$ with last vertex $v$ let $g^{h} := \langle (V,E), v, A, \{V_a\}_{a \in A}, (\prec^{h}_a)_{a \in A}\rangle$ be called the future game of $g$ after $h$, where for all $\rho, \rho' \in [\mathcal{H}_{g^h}]$ we set $\rho \prec^{h}_a \rho'$ iff $h\hat{\,}\rho \prec_a h\hat{\,}\rho'$. If $s$ is a strategy in $g$, let $s^{h}$ be the strategy in $g^{h}$ such that $s^{h}(h') := s(h\hat{\,}h')$ for all $h'\in \mathcal{H}_{g^h}$.
\end{itemize}
\end{definition}

\begin{observation}\label{obs:strat-derived-game}
Let $g = \langle (V,E), v_0, A, \{V_a\}_{a \in A}, (\prec_a)_{a \in A}\rangle$ be a game played on a finite graph.
\begin{enumerate}
\item\label{obs:strat-derived-game1} $g$ and its thresholds games have the same strategies.
\item\label{obs:strat-derived-game2} for all $h,h'\in \mathcal{H}$ ending with the same vertex the games $g^{h}$ and $g^{h'}$ have the same (finite-memory)
strategies.
\item\label{obs:strat-derived-game3} $g$, its future games, and their thresholds games have the same strategic implementations.
\item\label{obs:strat-derived-game4} If a strategy $s_a$ in $g$ is finite-memory, for all $h\in \mathcal{H}$ the strategy $s_a^{h}$ is also finite-memory.
\end{enumerate}
\begin{proof}
We only prove the fourth claim. Since $s_a$ is a finite-memory strategy, it comes from some strategic implementation $(\sigma,M_\epsilon)$. We argue that $(\sigma,M_{\sigma}(M_\epsilon,h)))$ is a strategic implementation for $s_a^{hv}$: First, $s_a^{hv}(v) = s_a(hv) = \pi_1\circ\sigma(v,M_{\sigma}(M_\epsilon,h)) = \pi_1\circ\sigma(v,M_{\sigma}(M_{\sigma}(M_\epsilon,h),\epsilon))$; second, for all $h'v'\in \mathcal{H}^{hv}$ we have $s_a^{hv}(vh'v') = s_a(hvh'v') = \pi_1\circ\sigma(v',M_{\sigma}(M_\epsilon,hvh')) = \pi_1\circ\sigma(v',M_{\sigma}(M_{\sigma}(M_\epsilon,h),vh'))$.
\end{proof}
\end{observation}

Our (transfer) results rely on players having winning strategies that are implementable with \emph{uniformly} finite memory, so that for every game they may be picked from a finite set of strategies. The following (shortenable) shorthands will be useful. Let $g$ be a game, let $a$ be a player.
\begin{itemize}
\item Let $m \in \mathbb{N}$ be such that in all threshold games for $a$ in $g$, if player $a$ has a winning strategy, she has one that is implementable using $m$ bits of memory. Then we say that player $a$ wins her winnable threshold games in $g$ using uniformly finite memory $m$.

\item Let $m \in \mathbb{N}$ be such that all (future) threshold games for $a$ in $g$ have finite-memory winning strategies that are implementable using $m$ bits of memory. Then we say that the (future) threshold games for $a$ in $g$ are uniformly-finite-memory determined using $m$ bits.
\end{itemize}
Note that speaking about future games above is the more general statement, as we prefix some finite history, and the sufficient memory depends on neither the threshold nor the history (cf Example \ref{example:future}).

\paragraph{Strict weak orders:}

The concepts so far are well-defined for preferences that are arbitrary binary relations. However, our results rely on the preferences being strict weak orders, as defined below, and all the preferences in the remainder of this article are assumed to be strict weak orders.

\begin{definition}[Strict weak order]
 \label{def:strictweakorder}
Recall that a relation $\prec$ is called a \emph{strict weak order} if it satisfies:
\[\begin{array}{l@{\quad}l}
\forall x,\quad \neg(x\prec x)  \\
\forall x,y,z, \quad x\prec y \,\wedge\, y\prec z\,\Rightarrow\, x\prec z  \\
\forall x,y,z, \quad \neg(x\prec y) \,\wedge\, \neg(y\prec z)\,\Rightarrow\, \neg(x\prec z)
\end{array}\]
\end{definition}

Strict weak orders capture in particular the situation where each player cares only about a particular aspect of the run (e.g.~her associated personal payoff), and is indifferent between runs that coincide in this aspect but not others (e.g.~the runs with identical associated payoffs for her, but different payoffs for the other players). We will show in Subsection \ref{subsec:lexreachability} that considering strict weak orders is strictly more general than working with payoff functions only.

\paragraph{Guarantees}

Definition~\ref{defn:ag-IAC} below rephrases Definitions 2.3 and 2.5 from~\cite{leroux3}: Given a strategy of a player $a$, the guarantee consists of the compatible runs plus the runs that are, according to $\prec_a$, not worse than all the compatible runs. The guarantee is thus upper-closed w.r.t. $\prec_a$. The best guarantee is the intersection of all the guarantees, and is thus also upper-closed.

 \begin{definition}[Player (best) future guarantee]\label{defn:ag-IAC}
Let $g$ be the game\\$\langle (V,E), v_0, A, \{V_a\}_{a \in A}, (\prec_a)_{a \in A}\rangle$ and let $a\in A$. For all $h\in \mathcal{H}$ and strategies $s_a$ for $a$ in $g^{h}$ let $\gamma_{a}(h,s_a) := \{\rho\in [\mathcal{H}_{g^h}] \,\mid\,\exists \rho'\in [\mathcal{H}_{g^h}(s_a)],\, \neg (\rho \prec_a^h \rho')\}$ be the player future guarantee by $s_a$ in $g^{h}$. Let $\Gamma_{a}(h):=\bigcap_{s_a}\gamma_a( h,s_a)$ be the best future guarantee of $a$ in $g^h$. We write $\gamma_a(s_a)$ and $\Gamma_a$ when $h$ is the trivial history.
\end{definition}

Note that in general the best guarantee may be empty, but our assumptions will (indirectly) rule this out: they will imply that each player has indeed a strategy realizing her best guarantee, i.e.~a strategy $s_a$ with $\Gamma_a = \gamma_a(s_a)$.

In Example \ref{example:guarantee} we construct a Nash equilibrium by starting with a strategy profile where everyone is realizing their guarantee, and then adding punishments against any deviators. The idea behind this construction is one ingredient of our results.

\begin{figure}[ht]
\centering
\begin{tikzpicture}[shorten >=1pt,node distance=1.6cm, auto]
  \node[state, initial] (x) {a};
  \node[draw, diamond] (y)  [right of = x] {b};
 \node[state] (x2) [below of = x] {x};
   \node[draw, diamond] (y2)  [below of = y] {y};

\path[->] (x) edge [bend left] node {} (y)
                edge node {} (x2)
                (x2) edge [loop left] node {} ()
                (y) edge [bend left] node {} (x)
                edge node {} (y2)
                (y2) edge [loop right] node {} ()       ;
 \end{tikzpicture}
  \caption{The game for Example \ref{example:guarantee}}
 \label{fig:fig1a}
 \end{figure}
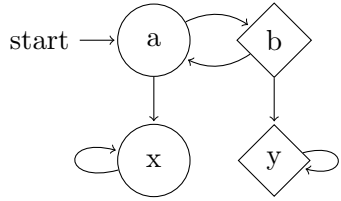

\begin{example}
\label{example:guarantee}
Let the underlying graph be as in Figure~\ref{fig:fig1a}, where circle vertices are controlled by Player 1 and diamond vertices are controlled by Player 2. The preference relation of Player $1$ is $(ab)^\omega \succ_1 a(ba)^nx^\omega \succ_1 (ab)^ny^\omega$ and the preference relation of Player $2$ is $(ab)^\omega \succ_2 (ab)^ny^\omega \succ_2 a(ba)^{n}x^\omega$ (in particular, both players care only about the tail of the run).

Then $\Gamma_1(a) = \{(ab)^\omega\} \cup \{a(ba)^nx^\omega \mid n \in \mathbb{N}\}$ and $\Gamma_2(a) = [\mathcal{H}]$. Player $1$ realizing her guarantee means for her to move to $x$ immediately, thus forgoing any chance of realizing the run $(ab)^\omega$. The Nash equilibrium constructed in the proof of both Theorem \ref{thm:fmNE-many-games} and Theorem \ref{thm:fmNE-one-game-bit} will be Player $1$ moving to $x$ and Player $2$ moving to $y$. Note that in this particular game, the preference of Player $2$ has no impact at all on the Nash equilibrium that will be constructed.
\end{example}

Note that the notion of best guarantee for a player does not at all depend on the preferences of the other players; and as such, it is a bit strenuous to consider a strategy realizing the guarantee (or the minimal runs therein) to be optimal in the game at hand (cf.~Example \ref{example:guarantee}). This strategy is rather optimal for a player that would play against a coalition of antagonistic opponents, and thus makes sense based on a worst-case assumption about the behaviour of the other players.

\paragraph{Optimality}

Strategies may be optimal (in a weak sense as discussed above) either at the beginning of the game or at all histories. This useful game-theoretic concept is rephrased below in terms of best guarantee.

\begin{definition}
\begin{itemize} Let $s_a$ be a strategy for some player $a$ in some game $g$.
\item $s_a$ is \emph{optimal}, if $\gamma_a(s_a) = \Gamma_a$.
\item $s_a$ is \emph{optimal at history} $h \in \mathcal{H}$, if $s_a^h$ is optimal in $g^h$.
\item $s_a$ is \emph{subgame-optimal}, if it is optimal at all $h \in \mathcal{H}$.
\item $s_a$ is \emph{consistent-optimal}, if it is optimal at all $h \in \mathcal{H}(s_a)$.
\end{itemize}
\end{definition}

Note that the notion of optimality is orthogonal to that of determinacy. Especially, the players having optimal strategies does not imply determinacy of the derived threshold games (in an undetermined win/lose game, the guarantee of both players is the set of all runs -- hence, any strategy is optimal).

The key Lemma \ref{lem:ufm-opt} essentially consists in a quantifier inversion, like the one in \cite{paulyleroux2}. Assuming that for all histories there is a finite-memory optimal strategy, we will use them to construct a finite-memory strategy that is subgame-optimal. To know when to use which strategy, we will use the assumption that optimality of a given strategy at an arbitrary history can be decided regularly.

\begin{definition}
A player $a$ in a game $g$ has the \emph{optimality is regular} (OIR) property, if for all finite-memory strategies $s_a$ for $a$ in $g$, there exists a finite automaton that decides on input $h\in \mathcal{H}$ whether or not $s_a$ is optimal at $h$.

A game $g$ has the OIR property, if all the players have it in $g$.

A preference has the optimality is regular property, if players with this preference have the OIR in all games.
\end{definition}

\section{Main results}
\label{sec:mainresults}

In the remainder of the paper, the games are always played on finite graphs has defined in Section~\ref{sec:background}, and they always involve colors in $C$, players in $A$, and preferences $(\prec_a)_{a\in A}$.

Theorem~\ref{thm:fmNE-many-games} presents implications, and absence of stated implication is discussed afterwards. For several implications we use the assumption that the set of preferences is closed under antagonism, \textit{i.e.} for all $\prec_a$ there exists $b \in A$ such that $\mathalpha{\prec}_b = \mathalpha{\prec}_a^{-1}$.

\begin{theorem}\label{thm:fmNE-many-games}
Let $(\prec_a)_{a\in A}$ be closed under antagonism. The statements below refer to all the games built with $C$, $A$, and $(\prec_a)_{a \in A}$, and the diagram displays implications between the statements.
\begin{itemize}
\item {\bf OIR:} Optimality is regular.
\item {\bf fm-SOS:} There are finite-memory subgame-optimal strategies.
\item {\bf fm-NE:} There are finite-memory Nash equilibrium.
\item {\bf FTG-d:} The future threshold games are determined.
\item {\bf TG-ufmd} In every game, the threshold games are determined using uniformly finite memory.
\item {\bf FTG-ufmd:} In every game, the future threshold games are determined using uniformly finite memory.
\end{itemize}
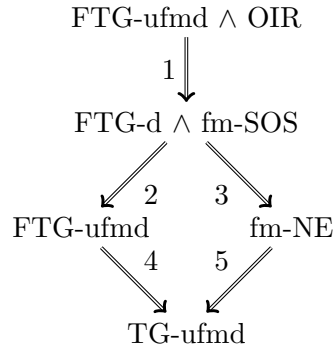
\centering
\begin{tikzpicture}[node distance=14mm]
\node(oir) {FTG-ufmd $\wedge$ OIR};
\node(sos) [below of = oir]{FTG-d $\wedge$ fm-SOS};
\node(middle) [below of = sos]{};
\node(ftg) [left of = middle]{FTG-ufmd};
\node(ne) [right of = middle]{fm-NE};
\node(tg) [below of = middle]{TG-ufmd};

\draw[->,double] (oir) -- node[left]{1} (sos);
\draw[->,double] (sos) -- node[below right]{2} (ftg);
\draw[->,double] (sos) -- node[below left]{3} (ne);
\draw[->,double] (ftg) -- node[above right]{4} (tg);
\draw[->,double] (ne) -- node[above left]{5} (tg);
\end{tikzpicture}

\begin{proof}
\begin{enumerate}
\item By Lemma \ref{lem:ufm-opt}.
\item Subgame-optimal strategies in an antagonistic two-player win all future threshold games that are winnable by their player.
\item By Lemma \ref{lemma:constructNE-all-th}.
\item Clear.
\item By Lemma \ref{lemma:NEuniformlyfmd}.
\end{enumerate}
\end{proof}
\end{theorem}

We will show in Subsection \ref{subsec:oir} that neither Implication (1)  (Example \ref{ex:not-oir2}) nor Implication (2) (Example \ref{ex:not-oir}) are reversible. For the remaining implications, we do not have answers regarding their reversibility. Regarding Implication (4), note that if we were to fix a specific graph, it would be trivial to separate the two notions. The difficulty lies in the requirement to deal with all finite graphs simultaneously.

\begin{question}
Is there a preference $\prec$ such that all threshold games for $\prec$ are uniformly finite memory determined, but not all future threshold games?
\end{question}

Regarding Implications (3) and (5), the ultimate goal would be precise characterization of the properties of (future) threshold games sufficient and necessary to have finite-memory Nash equilibria, akin to the characterization of the existence of optimal positional strategies by \name{Gimbert} and \name{Zielonka} \cite{gimbert}.

\begin{question}
How much more requirements are needed in addition to {\bf TG-ufmd} to imply {\bf fm-NE}?
\end{question}

Theorem~\ref{thm:fmNE-many-games} relies on many games having the assumed property, \textit{i.e} the statements are universal quantifications over games. However,  a given game might enjoy some property, \textit{e.g.} existence of NE, not only due to the preferences on infinite sequences of colors, but also due to the specific structure of the underlying graph. Theorem~\ref{thm:fmNE-one-game-bit} below captures this idea. It is a generalization of our main result in \cite{paulyleroux4-sr}. (The new Assumption~\ref{thm:fmNE-one-game-bit2} is indeed weaker.)

\begin{definition}
Given a game $g$, a preference $\prec$ is Mont (in $g$) if  for every regular run $h_0 \hat{\,}\rho\in [\mathcal{H}]$ and for every family $(h_n)_{n\in\mathbb{N}}$ of paths in $(V,E)$ such that $h_0 \hat{\,}\dots  \hat{\,}h_n \hat{\,}\rho \in [\mathcal{H}]$ for all $n$, if $h_0 \hat{\,}\dots  \hat{\,}h_n \hat{\,}\rho \prec h_0 \hat{\,}\dots  \hat{\,}h_{n+1} \hat{\,}\rho$ for all $n$ then $h_0\hat{\,}\rho \prec h_0 \hat{\,}h_1 \hat{\,}h_2 \hat{\,}h_3\dots$.
\end{definition}

\begin{theorem}\label{thm:fmNE-one-game-bit}
Let $g$ be a game such that
\begin{enumerate}
\item\label{thm:fmNE-one-game-bit1} For all $a\in A$ the future threshold games for $a$ in $g$ are determined, and each of Player $1$ and $2$ can win their winnable games via $k$ fixed strategies using $T$ bits each.

\item\label{thm:fmNE-one-game-bit2} Optimality for such a strategy is regular using $D$ bits.
\item The preferences in $g$ are Mont.
\end{enumerate}
Then $g$ has a Nash equilibrium made of strategies using $|A|(kD + kT + \log k) + 1$ bits each.
\begin{proof}
By Lemmata~\ref{lem:ufm-opt} and \ref{lemma:constructNE-ft}.
\end{proof}
 \end{theorem}

The difference between Theorem~\ref{thm:fmNE-one-game-bit} and Theorem~\ref{thm:fmNE-many-games} is similar to the difference between previous works by the authors for games on infinite trees (without memory concern): Namely, between \cite{leroux3} and \cite{paulyleroux2}. The first result characterizes the preferences that guarantee existence of NE for all games; the second result relies on the specific structure of a given game to guarantee existence of NE for the given game.

\section{Main proofs}
\label{sec:proofs}
This section organizes the lemmata for the two main results into several paragraphs, depending on the increasing strength of the assumptions.

\paragraph{Lemma without specific extra assumptions:} Lemma~\ref{lem:guarantee-inclusion} below collects basic useful facts on how the guarantees behaves wrt strategies and future games.

\begin{lemma}\label{lem:guarantee-inclusion}
Let $g$ be a game on a graph, let $\prec_a$ be a strict weak order preference for some player $a$, let $h\in \mathcal{H}$, let $s_a$ be a strategy for $a$ in $g^{h}$, let $h' \in \mathcal{H}(s_a)$, and let $s'_a$ be a strategy for $a$ in $g^{h\hat{\,}h'}$.
\begin{enumerate}

\item\label{lem:guarantee-inclusion2} Then $h'\hat{\,}\gamma_a(h\hat{\,}h',s_a^{h'}) \subseteq \gamma_a(h,s_a)$ for all $h'\in \mathcal{H}(s_a)$.\\

\item\label{lem:guarantee-inclusion5} If $\gamma_a(h\hat{\,}h',s'_a) \subsetneq \gamma_a(h\hat{\,}h', s_a^{h'})$, there exists $\rho \in \gamma_a(h\hat{\,}h',s_a^{h'})$ such that $\rho \prec_a^{h\hat{\,}h'} \rho'$ for all $\rho' \in \gamma_a(h\hat{\,}h',s'_a)$.

\item\label{lem:guarantee-inclusion6} If $\gamma_a(h\hat{\,}h',s'_a) \subseteq \gamma_a(h\hat{\,}h', s_a^{h'})$ then $h'\hat{\,}\gamma_a(h\hat{\,}h',s'_a) \subseteq \gamma_a(h, s_a)$.

\end{enumerate}
\begin{proof}
\begin{enumerate}

\item Let $\rho \in \gamma_a(h\hat{\,}h',s_a^{h'})$, so by Definition~\ref{defn:ag-IAC} there exists $\rho' \in [\mathcal{H}(s_a^{h'})]$ such that $\neg(\rho \prec_a^{h\hat{\,}h'} \rho')$, \textit{i.e.} $\neg(h'\hat{\,}\rho \prec_a^{h} h'\hat{\,}\rho')$. So $h'\hat{\,}\rho \in \gamma_a(h,s_a)$ since $h'\hat{\,}\rho' \in [\mathcal{H}(s_a)] \subseteq \gamma_a(h,s_a)$.

\item Let $\rho \in  \gamma_a(h\hat{\,}h', s_a^{h'}) \setminus \gamma_a(h\hat{\,}h',s'_a)$, so $\rho \prec_a^{h\hat{\,}h'}\rho'$ for all $\rho' \in \gamma_a(h\hat{\,}h',s'_a)$ by Definition~\ref{defn:ag-IAC}.

\item Let $\rho \in \gamma_a(h\hat{\,}h',s'_a)$, so $h'\hat{\,}\rho \in h'\hat{\,}\gamma_a(h\hat{\,}h', s_a^{h'})$ by assumption, so $h'\hat{\,}\rho \in \gamma_a(h, s_a)$ by Lemma~\ref{lem:guarantee-inclusion}.\ref{lem:guarantee-inclusion2}.

\end{enumerate}
\end{proof}
\end{lemma}

\paragraph{Using uniformly finite memory in the threshold games of a \emph{given} game:}

Lemma~\ref{lem:guarantee-th-ufmd} below establishes an equivalence between existence of simple optimal strategies and simple ways of winning threshold games, for a given player in a given game. Note that determinacy is not assumed.

\begin{lemma}
\label{lem:guarantee-th-ufmd}
Let $F_a$ be a finite set of strategies for some player $a$ in some game $g$. The following are equivalent.
\begin{enumerate}
\item\label{lem:guarantee-th-ufmd1} $a$ has an optimal strategy in $F_a$ for $g$;
\item\label{lem:guarantee-th-ufmd2} $a$ wins each of her winnable threshold games in $g$ via strategies in $F_a$;
\item\label{lem:guarantee-th-ufmd3} $a$ wins each of her winnable non-strict threshold games in $g$ via strategies in $F_a$.
\end{enumerate}
Moreover replacing "in $F_a$" above with "using $m$ bits of memory" is correct.
\begin{proof}
Let us assume \ref{lem:guarantee-th-ufmd1}, so let $s_a$ be such that $\gamma_a(s_a) = \Gamma_a$. By definition of $\Gamma_a$ there is no $s'_a$ such that $\gamma_a(s_a) \subsetneq \Gamma_a$, so $s_a$ wins all winnable (non-strict) threshold games in $g$, hence \ref{lem:guarantee-th-ufmd2} and \ref{lem:guarantee-th-ufmd3}.

Conversely let us assume \ref{lem:guarantee-th-ufmd2} or \ref{lem:guarantee-th-ufmd3}. For all $\rho \in [\mathcal{H}] \setminus \Gamma_a$, Player $1$ has a winning strategy $s_a^\rho \in F_a$ in the (non-strict) threshold game for $a$ and $\rho$ in $g$. In this case let $s_a^\rho$ be a winning strategy in $F_a$. By finiteness at least one of the $s_a^\rho$, which we name $s_a$, wins the (non-strict) threshold games for all $\rho \notin \Gamma_a$. This shows that $\gamma_a(s_a) \subseteq \Gamma_a$, so equality holds, hence optimality of $s_a$.

Moreover,  a player in a game with $n$ vertices has at most $(n2^{m})^{(n2^m)}$ strategies using $m$ bits of memory (by the $\sigma$ representation).
\end{proof}
\end{lemma}

In Lemma~\ref{lem:guarantee-th-ufmd} above, the implication $ \ref{lem:guarantee-th-ufmd1}.\,\Rightarrow  \ref{lem:guarantee-th-ufmd2}.\,\wedge\, \ref{lem:guarantee-th-ufmd3}.$ holds in a more general context. However, the finiteness of $F_a$ is key to the converse, as well as to the convenient remark that we can sometimes safely remain vague about whether we speak about strict or non-strict thresholds. This and Lemma~\ref{lem:guarantee-th-ufmd} are used in Lemma~\ref{lemma:NEuniformlyfmd} for antagonistic games below.

\begin{lemma}
\label{lemma:NEuniformlyfmd}
Let $F$ be a finite set of strategies for players $a$ and/or $b$ in some antagonistic game $g$. The following are equivalent.
\begin{enumerate}
\item\label{lemma:NEuniformlyfmd1} $g$ has an NE using strategies in $F$;
\item\label{lemma:NEuniformlyfmd2} the threshold games for $a$ in $g$ are determined via strategies in $F$;
\item\label{lemma:NEuniformlyfmd3} the threshold games for $a$ in $g$ are determined, and each player wins each of her winnable (non-strict) threshold games in $g$ via strategies in $F$.
\end{enumerate}
Moreover replacing "in $F$" above with "using $m$ bits of memory" is correct.
\begin{proof}
By Lemma~\ref{lem:guarantee-th-ufmd} the following are equivalent for all propositions $D$.
\begin{itemize}
\item $D\,\wedge$ each player has an optimal strategy for $g$ in $F$;
\item $D\,\wedge$ $a$ wins each of her winnable threshold games in $g$, and $b$ wins each of her winnable non-strict threshold games in $g$, all via strategies in $F$;
\item $D\,\wedge$ each player wins each of her winnable (non-strict) threshold games in $g$  via strategies in $F$.
\end{itemize}
Let $D$ be "the threshold games for $a$ in $g$ are determined", and let us prove that the above assertions correspond to \ref{lemma:NEuniformlyfmd1}, \ref{lemma:NEuniformlyfmd2}, and \ref{lemma:NEuniformlyfmd3} in the same order. It is a copy-paste for \ref{lemma:NEuniformlyfmd3}, it is a definition unfolding for \ref{lemma:NEuniformlyfmd2}, so let us focus on \ref{lemma:NEuniformlyfmd1}.

Let $s$ be an NE that uses strategies in $F$. Let $\rho$ be he run induced by $s$, and let $\rho_t$ be a threshold. If $\rho_t \prec_a \rho$, Player $1$ wins $g_{a,\rho_t}$ by using $s_a$; otherwise Player $2$ wins by using $s_b$. It shows that $D$ holds. Since $\rho$ is a $\prec_a$-minimum of $\gamma_a(s_a)$, the existence of some $s'_a$ such that $\gamma_a(s'_a) \subsetneq \gamma_a(s_a)$ would contradict $s$ being an NE, so $\gamma_a(s_a) = \Gamma_a$. And likewise $\gamma_b(s_b) = \Gamma_b$.

Conversely let us assume $D$ and let $s_a \in F$ and $s_b \in F$ satisfy $\gamma_a(s_a) = \Gamma_a$ and $\gamma_b(s_b) = \Gamma_b$. First note that $\Gamma_a \cap \Gamma_b$ is non-empty, as witnessed by the run $\rho$ induced by $(s_a,s_b)$. Since $\rho \in \Gamma_a$, Player $1$ cannot win $g_{a,\rho}$, so Player $2$ wins it, which implies that runs $\prec_a$-greater than $\rho$ are not in $\Gamma_b$. Likewise runs $\prec_b$-greater than $\rho$ are not in $\Gamma_a$. Therefore elements of $\Gamma_a \cap \Gamma_b$ are $\prec_a$-equivalent, and $(s_a,s_b)$ is an NE.
\end{proof}
\end{lemma}

Note that Lemmata~\ref{lem:guarantee-th-ufmd} and \ref{lemma:NEuniformlyfmd} can be extended to games in normal form since the two proofs do not use the sequentiality of the game at all.

\begin{definition}
If the assertions of Lemma~\ref{lemma:NEuniformlyfmd} hold, we say that the game is finite-memory determined, and its value is the equivalence class $\Gamma_a \cap \Gamma_b$.
\end{definition}

\begin{corollary}\label{cor:value-reg-witness}
If an antagonistic game is finite-memory determined, its value has a regular witness.
\begin{proof}
By Lemma~\ref{lemma:NEuniformlyfmd} there is a finite-memory NE, whose induced run is regular.
\end{proof}
\end{corollary}

\paragraph{Using uniformly finite memory in \emph{several} threshold games:}

Lemma~\ref{lemma:bestresponse1}.\ref{lemma:bestresponse11} below relies on Lemma~\ref{lem:guarantee-th-ufmd} to give a rather weak sufficient condition for a given player to have finite-memory best responses in all antagonistic games. Lemma~\ref{lemma:bestresponse1}.\ref{lemma:bestresponse12} relies on Lemma~\ref{lemma:bestresponse1}.\ref{lemma:bestresponse11} and will help us prove subgame-optimality in Lemma~\ref{lem:ufm-opt}, by allowing us to restrict our attention to regular runs. For this, $\textrm{reg}$ denotes the infinite sequences of the form $u_1\dots u_n (w_1\dots w_k)^\omega$.

\begin{lemma}\label{lemma:bestresponse1}
Fix a player $a$ and her preference such that for all one-player games $g_a$, player $a$ wins her winnable threshold games in $g_a$ using uniformly finite memory. Then in all antagonistic games involving $a$ and $b$,
\begin{enumerate}
\item\label{lemma:bestresponse11} all finite-memory strategies of $b$ is met with a finite-memory best response by $a$;
\item\label{lemma:bestresponse12} for all finite-memory strategies $s_b$, we have $\textrm{reg} \cap \gamma_b(s_b) = \textrm{reg} \cap \Gamma_b$ implies $\gamma_b(s_b) = \Gamma_b$.
\end{enumerate}

\begin{proof}
\begin{enumerate}
\item Let $g$ be such a game involving $a$ and $b$, and let $s_b$ be a finite-memory strategy of $b$, implemented by $\sigma_b$ using $m$ bits of memory. Let $g'$ be a defined as follows: the vertices are in $V' := V \times \{0,1\}^m$, for the coloring let $\Lambda'(v,q) := \Lambda(v)$, the preferences are like in $g$, and $((v,q)(v',q')) \in E'$ iff $(v,v') \in E \wedge q' = \pi_2\circ\sigma(v,q) \wedge (v \in V_b \Rightarrow v' = \pi_1\circ\sigma(v,q)))$. Therefore only player $a$ is playing in $g'$, and she can induce exaclty the same color traces as in $g$ when player $b$ plays according to $s_b$. By uniformity assumption and Lemma~\ref{lem:guarantee-th-ufmd} player $a$ has a finite-memory optimal strategy $s'_a$ in $g'$. Using $s'_a$ and $s_b$ one can construct a finite-memory best-response for $a$ to $s_b$ in $g$: player $a$ feeds the history $h$ in $g$ to $s_b$ and thus keeps track of the corresponding history $h'$ in $g'$. Player $a$ then uses $s'_a$ to compute a move in $g'$, which corresponds to a unique possible move to be played in $g$.

\item $\Gamma_b \subseteq \gamma_b(s_b)$ by definition. By Lemma~\ref{lemma:bestresponse1}.\ref{lemma:bestresponse11} let $s_a$ be a finite-memory best response to $s_b$. The run $\rho$ induced by $(s_a,s_b)$ is therefore a regular $\prec_b$-minimum of $\gamma_b(s_b)$. Since $\textrm{reg} \cap \gamma_b(s_b) = \textrm{reg} \cap \Gamma_b$, it is also in $\Gamma_b$.
\end{enumerate}
\end{proof}
\end{lemma}

The regular-Mont condition below is a weakening of the Mont condition, by considering only some regular runs.

\begin{definition}
A preference $\prec$ is regular-Mont if the following holds: for all $h_0, h_1, h_2, h_3 \in C^*$, if $h_0 h_1^n h_2h_3^\omega \prec h_0 h_1^{n+1} h_2h_3^\omega$ for all $n \in \mathbb{N}$, then $h_0 h_2h_3^\omega \prec h_0 h_1^\omega$.
\end{definition}

The contraposition of Lemma~\ref{lemma:regularmont} below will be used in the proof of Lemma~\ref{lem:ufm-opt} to say that suitable uniformity of a preference implies that it is regular-Mont.

\begin{lemma}
\label{lemma:regularmont}
Fix a player $a$ and her preference $\prec_a$. If $\prec_a$ is not regular-Mont, there is a one-player game where $a$ does not win her winnable (non-strict) threshold games with uniformly finite memory.
\begin{proof}
Let us assume that $\prec_a$ is not regular-Mont, so let $h_0,h_1,h_2, h_3 \in C^*$ be such that $h_0h_1^nh_2h_3^\omega \prec h_0h_1^{n+1}h_2h_3^\omega$ for all $n\in\mathbb{N}$, but $\neg(h_0h_2h_3^\omega \prec h_0h_1^\omega)$. So $h_0h_1^\omega \prec h_0h_1h_2h_3^\omega$, since $\prec_a$ is a strict weak order. In the game below, player $a$ can win all the thresholds $h_0h_1^nh_2h_3^\omega$ but requires unbounded finite memory.

\begin{tikzpicture}[shorten >=1pt,node distance=2cm, auto]
  \node[state, initial] (x) {};
  \node[state] (y)  [right of = x] {};
  \node[state] (z)  [right of = y] {};

\path[->]   (x) edge node {$h_0$} (y)
                (y) edge [loop above] node {$h_1$} ()
                (y) edge node {$h_2$} (z)
                (z) edge [loop above] node {$h_3$} ();
 \end{tikzpicture}
\end{proof}
\end{lemma}

Lemmata~\ref{lemma:constructNE-all-th} and \ref{lemma:constructNE-ft} both show the existence of finite-memory NE. They both assume existence of finite-memory consistent-optimal strategies, but the other assumptions are slightly different. Lemma~\ref{lemma:constructNE-all-th} will be weakened (for the sake of simplicity) to obtain Implication $3$ of Theorem~\ref{thm:fmNE-many-games}; whereas Lemma \ref{lemma:constructNE-ft}, making assumptions only about the given game and its derived games, will be combined to Lemma~\ref{lem:ufm-opt} to obtain Theorem~\ref{thm:fmNE-one-game-bit}. The proofs ot Lemmata~\ref{lemma:constructNE-all-th} and \ref{lemma:constructNE-ft} are similar: The beginning is the same and is writen only once; the ends are similar yet not enough to copy-paste it; and the middle parts are clearly different, thus making a factorization of the two lemmata difficult.

\begin{lemma}
\label{lemma:constructNE-all-th}
Let $g$ be a  game, and let us assume the following.
\begin{enumerate}
\item All players have consistent-optimal strategies in $g$ using $S$ bits.
\item For all players $a$, for all games with $n$ vertices, all (non-strict) threshold games for $a$ are determined and Player $2$ wins her winnable ones using $f(n)$ bits.
\end{enumerate}
Then $g$ has a Nash equilibrium made of strategies using $\max(|A|S, f(|V|(1+2^{|A|S}))) + 1$ bits each.
\begin{proof}
For all $a$ let $s_a$ be a consistent-optimal finite-memory strategy for $a$ in $g$, and let $\rho_{NE}$ be the run induced by the strategy profile $(s_a)_{a \in A}$. It will become the run induced by the claimed finite-memory NE, once we ensure via finite memory that no player has an incentive to deviate.

By finiteness of memory, $\rho_{NE} \in \textrm{reg}$, so let $l$ be a lasso arena that corresponds to $\rho_{NE}$, \textit{i.e.} where the copies of a vertex from $g$ are labeled with the same player and color. Note that the lasso may be chosen with at most $|V|2^{|A|S}$ vertices, since when the $\rho_{NE}$ comes back to a previously visited vertex with all the players having the same memory content, the lasso is cycling.

For every player $a$, we construct a game $g^a$ as follows: the preferences are as in $g$, we take the disjoint union of $l$ and the arena of $g$, and we let $g^a$ start at the start of $l$. Finally, there is an edge from $v_l \in l$ to $v \in g$ if $v_l$ is controlled by $a$ and if there is one edge in $g$ from $v'$ to $v$, where $v_l$ is a copy of $v'$. So the players other than $a$ cannot deviate from $l$.

$g^a_{a,\rho_{NE}}$ is the threshold game derived from $g^a$ for player $a$ and threshold $\rho_{NE}$. Since $s_a$ is consistent-subgame optimal, it is optimal at all finite prefixes of $\rho_{NE}$, \textit{i.e.} $\rho \in \Gamma_a(h)$ for all decompositions $\rho_{NE} = h \hat{\,} \rho$, so Player $1$ loses $g^a_{a,\rho_{NE}}$  (on behalf of player $a$). This game is finite-memory determined by assumption, so let $s_{-a}$ be a finite-memory winning strategy for Player $2$.

Now let $s'_a$ be the following strategy for player $a$ in $g$: Follow $s_a$ until a player $b$ deviates, in which case play anything positionally if $b = a$, and follow $s_{-b}$ otherwise. This ensures that $b$ cannot get a better outcome by deviating unilaterally, so $(s'_a)_{a\in A}$ is an NE.

Every player uses $S$ bits of memory to follow $\rho_{NE}$, and she uses $(|A|-1)S$ bits to know how the others are supposed to play and thus detect when someone has deviated and who. To be able to take part in the coalition "punishing" the deviator, she uses $f(|V|(1+2^{|A|S}))$ bits to remember the $s_{-a}$. Since the two phases of the play, \textit{i.e.} following $\rho_{NE}$ and "punishing" a deviator, are not simultaneous, the memory can be repurposed: altogether $\max(|A|S, f(|V|(1+2^{|A|S}))) + 1$ bits suffice, where the $+1$ is used to remember the current phase.
\end{proof}
\end{lemma}

\paragraph{Using uniformly finite memory for the \emph{future} threshold games of a \emph{given} game:}

\begin{lemma}
\label{lemma:constructNE-ft}
Let $g$ be a game  satisfying the following.
\begin{enumerate}
\item All players have consistent-optimal strategies in $g$ using $S$ bits.
\item\label{lemma:constructNE-ft2} For each player, her future threshold games in $g$ are determined, and Player $2$ wins her winnable ones using $k$ fixed strategies using $T$ bits each.
\item Optimality for each strategy from Assumption~\ref{lemma:constructNE-ft2} is regular using $D$ bits.
\end{enumerate}
Then $g$ has a Nash equilibrium made of strategies using $|A|\max(S, kD + kT) + 1$ bits each.
\begin{proof}
Let $s_a$ and $\rho_{NE}$ be as in the proof of Lemma~\ref{lemma:constructNE-all-th}, and likewise let us build a strategy $s_{-a}$ that prevents $a$ from deviating. For all $a \in A$ let $t_{-a}^1,\dots,t_{-a}^k$ be the strategies from Assumption~\ref{lemma:constructNE-ft2}, and for each $t_{-a}^i$ let $A_i$ be an automaton using $D$ bits of memory and telling for which histories $t_{-a}^i$ is optimal. Let $s'_a$ be the following strategy for player $a$ in $g$: Follow $s_a$ until a player $b$ deviates, in which case play anything positionally if $b = a$, and otherwise take part in the optimal $t_{-b}^i$ of smallest index $i$.

Consider the future games starting right after deviation of $b$. Their antagonisitic versions for $a$ against the others have values at most $\rho_{NE}$ wrt $\prec_b$, by optimality of $s_b$. So choosing the right $t_{-b}^i$ as above ensures that $b$ cannot get a better outcome by deviating unilaterally from $\rho_{NE}$, so $(s'_a)_{a\in A}$ is an NE.


Every player uses $S$ bits of memory to follow $\rho_{NE}$, and she uses $(|A|-1)S$ bits to know how the others are supposed to play and thus detect when someone has deviated and who. To be able to take part in the coalition "punishing" the deviator, she uses $(|A|-1)kT$ bits to remember the $t_{-a}^i$ for all $a$ but herself; and she uses $(|A|-1)kD$ bits to know when a $t_{-a}^i$ is optimal. Since the two phases of the play, \textit{i.e.} following $\rho_{NE}$ and "punishing" a deviator, are not simultaneous, the memory can be repurposed: altogether $|A|\max(S, kD + kT) +1$ bits suffice, where the $+1$ is used to remember the current phase.
\end{proof}
\end{lemma}

\paragraph{Using uniformly finite memory for \emph{several} \emph{future} threshold games:}

Lemma~\ref{lemma:bestresponse2} is a complex variant of Lemma~\ref{lemma:bestresponse1}.\ref{lemma:bestresponse11} that considers future games. It is not invoked in our main results but will be useful in Subsection~\ref{subsec:oir} to further weaken the assumption that optimality is regular.

\begin{lemma}
\label{lemma:bestresponse2}
Fix a player $a$ and her preference such that for all one-player games $g_a$, player $a$ wins her winnable future threshold games in $g_a$ via uniformly finite memory. Then in all antagonistic games involving $a$ and $b$, for all finite memory strategies $s_b$, there are finitely many finite-memory strategies $t_1, \ldots, t_n$ such that for all histories $h$, one of the $t_i$ is a best response by $a$ to $s_b$ starting at $h$.
\begin{proof}
(This proof is similar to that of Lemma~\ref{lemma:bestresponse1}.\ref{lemma:bestresponse11}.) Let $g$ be such a game involving $a$ and $b$, and let $s_b$ be a finite-memory strategy of $b$. Let $g'$ be defined wrt $g$ as in the proof of Lemma~\ref{lemma:bestresponse1}.\ref{lemma:bestresponse11}. Only player $a$ is playing in $g'$, so by uniformity assumption let finite-memory strategies $t'_1, \ldots, t'_n$ be such that all winnable future threshold games in $g'$ is won by some $t'_i$. Using $t'_i$ and $s_b$ one can construct a finite-memory strategy $t_i$ : player $a$ feeds the history $h$ in $g$ to $s_b$ and thus keeps track of the corresponding history $h'$ in $g'$. Player $a$ then uses $t'_i$ to compute a move in $g'$, which corresponds to a unique possible move to be played in $g$. This defines a strategy $t_i$ in $g$. Now for all $h$ in $g$, some $t'_i$ is an optimal strategy for $a$ at $h'$ in $g'$, and the corresponding $t_i$ is is a best response by $a$ to $s_b$ starting at $h$.
\end{proof}
\end{lemma}

\paragraph{Main construction:}

Lemma~\ref{lem:ufm-opt} below concludes that a player $a$ has a subgame optimal strategy. Assumptions~\ref{cond:ufm-opt1} and \ref{cond:ufm-opt2} suffice to construct the candidate strategy, whereas the other assumptions are only used to prove subgame optimality. Assumption~\ref{cond:ufm-opt3} is used to build uniformly finite memory best responses by $b$ for all histories. Then Assumption~\ref{cond:ufm-opt41} (resp. \ref{cond:ufm-opt42}) allows us to conclude quickly (resp. to continue the proof). Assumptions~\ref{cond:ufm-opt3} and \ref{cond:ufm-opt422} are partly redundant due to our factoring out of two theorems and proofs. Note that determinacy need not hold.

\begin{lemma}
\label{lem:ufm-opt}
 Let $a$ and $b$ be the players of an antagonistic game $g$. Let us assume the following:
 \begin{enumerate}
\item\label{cond:ufm-opt1} player $a$ wins her winnable future threshold games in $g$ via $k$ strategies using $T$ bits each,
\item\label{cond:ufm-opt2} Optimality of such a strategy is regular using $D$ bits,
\item\label{cond:ufm-opt3} players $b$ wins her winnable future threshold games either in $g$ or in every one-player game using uniformly finite memory.
\item Either of the following holds:
\begin{enumerate}
\item\label{cond:ufm-opt41}  $a$'s preference is Mont;
\item\label{cond:ufm-opt42}
\begin{enumerate}
\item\label{cond:ufm-opt421} optimality is regular for $b$'s strategies in $g$, and
\item\label{cond:ufm-opt422} Each of players $a$ and $b$ wins her winnable threshold games in every one-player game using uniformly finite memory.
\end{enumerate}
\end{enumerate}
\end{enumerate}
Then player $a$ has a subgame-optimal strategy in $g$ using $kD + kT + \log k$ bits.
\begin{proof}
 Let $t_1,\dots, t_k$ be the strategies from Assumption~\ref{cond:ufm-opt1}. By Lemma~\ref{lem:guarantee-th-ufmd}, for all $h$ one of the $t_i$ is optimal at $h$. By Assumption~\ref{cond:ufm-opt2} there exist automata $A_1,\dots, A_k$ using $D$ bits that decide their respective optimality depending on $h$.

We define a strategy $s$ for player $a$ as follows: always store the index of one of the $t_i$, and follow the selected $t_i$ until it ceases to be optimal at some history. Then select an optimal $t_j$ and follow it. Storing the index of the strategy requires $\log k$ bits; simulating all the $k$ strategies in parallel uses $kT$ bits; and deciding optimality $kD$ bits, so $kD + kT + \log k$ bits suffice to implement this strategy.

It remains to show that $s$ is indeed subgame-optimal.

To show that $\gamma_a(h, s^{h}) = \Gamma_a(h)$ for all $h$, let $h_0 \in \mathcal{H}$, let $\rho \in \mathcal{H}(s^{h_0})$. In a slight notation overload, for all histories $h$ let $t_h$ be the one strategy $t_i$ that our construction of $s$ follows at $h$. Also, let $(h_n)_{n\geq 1}$ be such that the $n$-th change occurring strictly after $h_0$ occurs at history $h'_n :=h_0 \hat{\,} h_1\hat{\,}\dots \hat{\,}h_n$, and let us make a case disjunction on whether $(h_n)_{n\geq 1}$ is finite or infinite. First case, player $a$ changes strategies finitely many times along $\rho$, say $N$ times. Applying Lemma~\ref{lem:guarantee-inclusion}.\ref{lem:guarantee-inclusion6} $N$ times yields $h_1\hat{\,}\dots \hat{\,}h_{N}\hat{\,}\gamma_a(h'_N,t_{h'_N}^{h'_N}) \subseteq \dots\subseteq  h_{1}\hat{\,}\gamma_a(h'_1,t_{h'_1}^{h'_1})\subseteq \gamma_a(h_0,t_{h_0}^{h_0}) \subseteq \Gamma_a(h_0)$. So $\rho \in \Gamma_a(h_0)$ since $\rho\in h_1\hat{\,}\dots\hat{\,}h_{N}\hat{\,}\gamma_a(h'_N,t_{h'_N}^{h'_N})$.

Second case, player $a$ changes strategies infinitely many times along $\rho$. Regardless of which disjunct of Assumption~\ref{cond:ufm-opt3} holds, there are finitely many finite-memory strategies $r_i$ for $b$ in $g$ such that for all $h$ one of the $r_i$, which we call $r_h$, is a best-response by $b$ to $t_h$. So $(r_h,t_h)$ induces a minimum $\rho_h$ of $\Gamma_a(h)$. Since the $t_i$ and the $r_i$ are finitely many and are finite-memory strategies, the $\rho_h$ are also finitely many, and regular.

Since $a$ changes strategies at $h'_{n+1}$ we have $h'_{n} \hat{\,} \rho_{h'_n} \prec_a h'_{n+1} \hat{\,} \rho_{h'_{n+1}}$ (or $\rho_{h'_n} \prec_a^{h'_{n}} h_{n+1} \hat{\,} \rho_{h'_{n+1}}$). By finiteness some $\rho_{h}$ must occur infinitely often, let $\rho'$ be the constant run of the the corresponding subsequence $\rho_{h'_{\varphi(n)}}$. So $h'_{\varphi(0)}  \hat{\,} \rho' \prec_a  \dots \prec_a h'_{\varphi(n)} \hat{\,} \rho' \prec_a \dots$. If Assumption~\ref{cond:ufm-opt41} holds, $a$'s preference is Mont and therefore $h_0  \hat{\,} \rho' \prec_a \rho$, \textit{i.e.} $\rho \in \Gamma_a(h_0)$ since $h_0  \hat{\,} \rho' \in \Gamma_a(h_0)$.

Let us now deal with the case where Assumption~\ref{cond:ufm-opt42} holds. By Lemma~\ref{lemma:bestresponse1}.\ref{lemma:bestresponse12} and Assumption~\ref{cond:ufm-opt422} (for $b$) we can assume wlog that $\rho$ is regular. By Assumption~\ref{cond:ufm-opt421}, suitable $r_h$ can also be chosen via a finite automaton, in which case $h'_{\varphi(n)}$ can be decomposed as $h_0uw^n$. So $h_0u \hat{\,} \rho' \prec_a h_0uw \hat{\,} \rho' \prec_a \dots \prec_a h_0uw^n \hat{\,} \rho' \prec_a \dots$. By Assumption~\ref{cond:ufm-opt422} (for $a$) and contraposition of Lemma~\ref{lemma:regularmont} we have $h_0 u \hat{\,} \rho' \prec_a \rho$, \textit{i.e.} $\rho \in \Gamma_a(h_0)$ since $h_0  \hat{\,} \rho' \in \Gamma_a(h_0)$.
\end{proof}
\end{lemma}

\section{Discussion}
\label{section:discussion}
\subsection{Comparison to previous work}
As mentioned above, a similar but weaker result (compared to our Lemma \ref{lemma:constructNE-ft}) has previously been obtained by \name{Brihaye}, \name{De Pril} and \name{Schewe} \cite{depril2},\cite[Theorem 4.4.14]{depril}. They use cost functions rather than preference relations. Our setting of strict weak orders is strictly more general \footnote{For example, the lexicographic combination of two payoff functions can typically not be modeled as a payoff function, as $\mathbb{R} \times \{0,1\}$ (with lexicographic order) does not embed into $\mathbb{R}$ as a linear order, cf.~Subsection \ref{subsec:lexreachability}.}\label{page:footnote}. However, even if both frameworks are available, it is more convenient for us to have results formulated via preference relations rather than cost functions: Cost functions can be translated immediately into preferences, whereas translating preferences to cost functions is more cumbersome. In particular, it can be unclear to what extend \emph{nice} preferences translate into \emph{nice} cost functions. Note also that prefix-linearity for strict weak orders is more general than prefix-linearity for cost functions. We will see in Subsection \ref{subsec:oir} that prefix-linearity implies the optimality-is-regular property by a very simple argument.

As a second substantial difference, \cite[Theorem 4.4.14]{depril} requires either prefix-independent cost functions and finite-memory determinacy of the induced threshold games, or prefix-linear cost functions and optimal positional strategies in the induced antagonistic games. In particular,\cite[Theorem 4.4.14]{depril} cannot be applied to bounded energy parity games, where finite prefixes of the run do impact the overall value for the players, and where at least the protagonist requires memory to execute a winning strategy.

Before \cite{depril,depril2}, it had already been stated by \name{Paul} and \name{Simon} \cite{soumya} that multi-player multi-outcome Muller games have Nash equilibria consisting of finite memory strategies. As (two-player win/lose) Muller games are finite-memory determined \cite{gurevich2}, and the corresponding preferences are obviously prefix independent, this result is also a consequence of \cite[Theorem 4.4.14]{depril}. Another result subsumed by \cite[Theorem 4.4.14]{depril} (and subsequently by our main theorem) is found in \cite{depril3} by \name{Brihaye}, \name{Bruy\`ere} and \name{De Pril}.

\subsection{Exploring optimality is regular}
\label{subsec:oir}
We shall discuss the optimality-is-regular property and its relationship to some other, established properties of preferences. We will in particular show that it is not dispensable in our main theorem (Example \ref{ex:not-oir}), as in its absence, uniform finite-memory determinacy no longer implies the existence of finite memory subgame-optimal strategies. On the other hand, the optimality-is-regular property is also not necessary, as shown by Example \ref{ex:not-oir2}.

Recall that a preference relation $\prec \subseteq [\mathcal{H}] \times [\mathcal{H}]$ is called \emph{prefix-linear}, if $\rho \prec \rho' \Leftrightarrow h \hat{\,}\rho \prec h \hat{\,} \rho'$ for all $\rho , \rho', h \hat{\,}\rho \in [\mathcal{H}]$. It is \emph{prefix-independent}, if $\rho \prec \rho' \Leftrightarrow h  \hat{\,}\rho \prec \rho'$ and $\rho' \prec \rho \Leftrightarrow \rho' \prec h \hat{\,}\rho $ for all $\rho , \rho', h \hat{\,}\rho \in [\mathcal{H}]$. Clearly, a prefix-independent preference is prefix-linear.

As a further generalization, we will consider \emph{automatic-piecewise prefix-linear} preferences $\prec$. Here, there is an equivalence relation on $\mathcal{H}$ with equivalence classes (pieces for short) in $\overline{\mathcal{H}}$ and satisfying three constraints: First, the histories in the same piece end with the same vertex. Second, there exists a deterministic finite automaton, without accepting states, that reads histories and  such that two histories are equivalent iff reading them leads to the same states. Third, for all $h \hat{\,}\rho , h \hat{\,}\rho', h' \hat{\,}\rho,h' \hat{\,}\rho' \in [\mathcal{H}]$, if $ \overline{h'} = \overline{h}\in \overline{\mathcal{H}}$, then $h \hat{\,}\rho \prec h  \hat{\,}\rho' \Leftrightarrow h'  \hat{\,}\rho \prec h'  \hat{\,}\rho'$.

The extension to automatic-piecewise prefix-linear preferences ensures that e.g.~safety and reachability games are also covered. In fact, most of the common payoff functions considered in the literature give rise to automatic-piecewise prefix-linear preferences. Examples include mean-payoff, discounted payoff, Muller, mean-payoff parity and bounded energy parity games (see Subsection \ref{subsec:energyparity} below for the latter).

Of the popular winning conditions, many are actually prefix-independent, such as parity, Muller, mean-payoff, cost-Parity, cost-Streett \cite{fijalkow2}, etc. Clearly, any combination of prefix-independent conditions itself will be prefix-independent. Typical examples of non-prefix independent, but prefix-linear conditions are reachability, energy, and discounted payoff. Combining prefix-linear conditions not necessarily yields another prefix-linear condition. However, we can easily verify that combining a reachability or energy condition with any prefix-linear condition yields an automatic-piecewise prefix-linear condition (provided that energy is bounded).


\begin{proposition}
\label{prop:applimpliesoir}
Automatic-piecewise prefix-linear preferences have the optimality-is-regular property.
\begin{proof}
Assume automatic-piecewise prefix-linear preferences. Whether a finite memory strategy is optimal at some history depends only on its memory content at that history, and on the piece the history falls into.
\end{proof}
\end{proposition}

\begin{proposition}
\label{prop:applcharac}
A preference $\prec$ is automatic-piecewise prefix-linear iff there is a finite set $\mathcal{M}$ of regular sets $M \subseteq \mathcal{H}$ such that:
\[ \forall p, q \in C^\omega \quad \exists M \in \mathcal{M} \quad \forall h \in \mathcal{H} \quad \left (h \in M \Leftrightarrow h\hat{\,}p \prec h\hat{\,}q \right )\]
\begin{proof}
If $\prec$ is automatic-piecewise prefix-linear, then the pieces satisfy the property of $\mathcal{M}$. Conversely, given some finite set $\mathcal{M}$ with the given property, we can consider the finest partition where each part is some boolean combination of elements of $\mathcal{M}$. This partition is automatic, and witnesses automatic-piecewise prefix linearity $\prec$.
\end{proof}
\end{proposition}

\begin{definition}
We say that $\prec$ has the \emph{weak oir} property if
\[ \forall p, q \in C^\omega \cap \textrm{reg} \quad \exists M \in  \textrm{Reg} \quad \forall h \in \mathcal{H} \quad \left (h \in M \Leftrightarrow h\hat{\,}p \prec h\hat{\,}q \right )\]
where $\textrm{Reg}$ are the regular subsets of $C^*$.
\end{definition}

\begin{observation}
The OIR property implies the weak OIR property.
\begin{proof}
Given regular $p$ and $q$, we can construct finite graphs $P$ and $Q$, such that the only infinite run through $P$ has colors $p$, and the only infinite run through $Q$ has colors $q$. We further use a clique $K$ where all colors appear, controlled by a different player. Then we merge them together as follows: The game starts in $K$, from where a choice vertex $v$ controlled by the protagonist can be reached. The vertex $v$ has two outgoing edges, to $P$ and to $Q$.

Now we use the OIR property on the constant strategy going to $P$. At some history $h$ ending in $v$, we find that $h\hat{\,}p \prec h\hat{\,}q$ iff this constant strategy is not optimal at $h$. By using the construction with all finitely many different choices for the color of $v$, we obtain the full claim.
\end{proof}
\end{observation}

\begin{proposition}\label{prop:weak-oir-oir}
Let $g$ be an antagonistic game involving $a$ and $b$, and let us assume the following.
\begin{enumerate}
\item\label{prop:weak-oir-oir1} for all one-player games $g_b$, player $b$ wins her winnable future threshold games in $g_b$ via uniformly finite memory.
\item\label{prop:weak-oir-oir2} Player $a$ wins her winnable future threshold games in $g$ using uniformly finite memory.
\item\label{prop:weak-oir-oir3} $\prec_a$ has the weak OIR property
\end{enumerate}
Then $\prec_a$ has also the OIR property.
\begin{proof}
Let $s_a$ be a finite-memory strategy for $a$ in some antagonistic game $g$ involving $a$ and $b$. We want to construct an automaton that decides whether $s_a$ is optimal at some input history. By Assumption~\ref{prop:weak-oir-oir1} and Lemma~\ref{lemma:bestresponse2}, there are finitely many finite-memory strategies $s_b^i$ such that for every history $h$, one of the $s_b^i$ is a best response by $b$ to $s_a$ at $h$. Depending on the $h$ from which we start, playing some fixed $s_b^i$ against $s_a$ yields a tail from some finitely many tails $p_{ij}$. Moreover we can decide which $j$ from $h$ (and $i$) in an automatic way.

By Assumption~\ref{prop:weak-oir-oir2} and Lemma~\ref{lem:guarantee-th-ufmd} there are finitely many finite-memory strategies $t_a^k$ for $a$ such that for all histories $h$, one of the $t_a^k$ is optimal at $h$. By Assumption~\ref{prop:weak-oir-oir1} and Lemma \ref{lemma:bestresponse2} again, there are finitely many finite-memory strategies $t_b^l$ such that for all histories $h$ and strategies $t_a^k$, one of the $t_b^l$ is a best response to $t_a^k$. Depending on the $h$ from which we start, playing some fixed $t_b^j$ against $t_a^i$ yields a tail from some finitely many tails $q_{klm}$. Moreover we can decide which $m$ from $h$ (and $k$, $l$) in an automatic way.

Now, $s_a$ is not optimal at some history $h$ iff one of the $t_a^i$ does better at $h$, which is equivalent to \[\min_{i} h\hat{\,}p_{ij(i,h)} \prec_a \max_{k} \min_{l} h\hat{\,}q_{klm(k,l,h)}\]
By invoking the weak OIR property for the finitely many tails $p_{ij}$ and $q_{klm}$, and using that $j$ and $m$ depend on $h$ in an automatic way, we can test this property with a finite automaton.
\end{proof}
\end{proposition}

\begin{corollary}
If all future threshold games are uniformly finite-memory determined, then $\prec$ has the weak OIR property iff it has the OIR property.
\end{corollary}

\begin{example}
\label{ex:not-oir}
Fix some non-regular set $A \subseteq \mathbb{N}$. We define a payoff function $P_A : \{0,1\}^\omega \to \mathbb{N} \cup \{+\infty\}$ as follows:
\[P_A(p) = \begin{cases} +\infty & p = 0^\omega\\
 2n+1 & \exists q \in \{0,1\}^\omega \ (p = 0^n10q \wedge n \in A) \vee (p = 0^n11q \wedge n \notin A)
\\ 2n & \exists q \in \{0,1\}^\omega \ (p = 0^n10q \wedge n \notin A) \vee (p = 0^n11q \wedge n \in A)
\end{cases}\]

The induced preference $\prec_A$ is given by $p \prec_A q$ iff $P_A(p) < P_A(q)$.

{\bf Claim}: The threshold games for $\prec_A$ are uniformly finite-memory determined.
\begin{proof}
If the protagonist can win the safety game for staying on vertices colored $0$, he has a positional strategy for doing so. This strategy would win all winnable threshold games on that graph. If he can not win that game, then the opponent can force a $1$ within $k \leq n$ moves (where $n$ is the size of the graph), and she can do so positionally. In particular, the protagonist loses all threshold games for thresholds $0^j11q$ or $0^j10q$ for $j > k$. As all threshold games here have $\omega$-regular winning conditions, and there are only finitely many cases left, uniform finite-memory determinacy follows.
\end{proof}

{\bf Claim}: The games built with $\prec_A$ do not have the optimality-is-regular property, and the player with preference $\prec_A$ does not always have a finite-memory subgame-optimal strategy.
\begin{proof}
Consider the game graph depicted in Figure \ref{fig:not-oir}, with the protagonist (with preference $\prec_A$) controlling the diamond vertex and the opponent the circle vertices. Further, consider the positional strategy where the protagonist always goes to the vertex labeled $0$. If there were an automaton that decides whether this strategy is optimal after some history, then by applying this automaton to histories of the form $0^n1$ allows us to decide whether $n \in A$, a contradiction to the choice of $A$ being non-regular. Similarly, by inspecting the choice a finite-memory subgame-optimal strategy of the player makes after some history of the form $0^n1$ allows us to decide whether $n \in A$, again a contradiction.
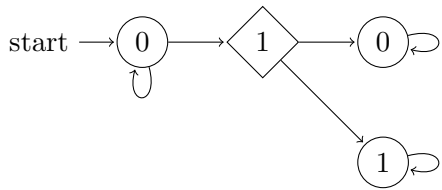
\begin{figure}[ht]
\centering
\begin{tikzpicture}[shorten >=1pt,node distance=1.6cm, auto]
  \node[draw, circle, initial] (x) {$0$};
   \node[draw, diamond] (z)  [right of = x] {$1$};
  \node[draw, circle] (a)  [right of = z] {$0$};
  \node[draw, circle] (b) [below of = a] {$1$};

\path[->]  (x) edge node {} (z)
                (x) edge [loop below] node {} ()
                (z) edge node {} (a)
                (z) edge node {} (b)
                (a) edge [loop right] node {} ()
                (b) edge [loop right] node {} ();
 \end{tikzpicture}
  \caption{The graph for the game in Examples \ref{ex:not-oir} and \ref{ex:not-oir2}}
 \label{fig:not-oir}
 \end{figure}
\end{proof}
\end{example}

\begin{example}
\label{ex:not-oir2}
Fix some non-regular set $B \subseteq \mathbb{N}$. We define a payoff function $P_B : \{0,1\}^\omega \to \mathbb{N} \cup \{-\infty,+\infty\}$ as follows:
\[P_B(p) = \begin{cases} +\infty & p = 0^\omega\\
 2n+1 & n \in B \wedge \exists q \in \{0,1\}^\omega \quad p = 0^n10q
\\ 2n & n \notin B \vee \exists q \in \{0,1\}^\omega \quad p = 0^n11q
\end{cases}\]

The induced preference $\prec_B$ is given by $p \prec_B q$ iff $P_B(p) < P_B(q)$.

{\bf Claim}: The player with preference $\prec_B$ always has a finite-memory subgame-optimal strategy.
\begin{proof}
The player first plays a safety game where he tries to stay on vertices colored $0$ as long as possible. If a $1$ is ever reached, and the player can choose, he will go a vertex colored $0$ in the next step.
\end{proof}

{\bf Claim}: The games built with $\prec_B$ do not have the optimality-is-regular property.
\begin{proof}
Consider the again game graph depicted in Figure \ref{fig:not-oir}, with the protagonist (with preference $\prec_B$) controlling the diamond vertex and the opponent the circle vertices. Further, consider the positional strategy where the protagonist always goes to the vertex labeled $1$. If there were an automaton that decides whether this strategy is optimal after some history, then by applying this automaton to histories of the form $0^n1$ allows us to decide whether $n \in B$, a contradiction to the choice of $B$ being non-regular.

\end{proof}
\end{example}

\subsection{On the Mont condition}
If we consider \emph{all} games on finite graphs involving a certain set of preferences, we saw by Lemma \ref{lemma:regularmont} that the regular-Mont condition comes for free in our setting. If we explicitly assume the regular-Mont condition to hold, it suffices on the other hand to make assumptions merely about games played on some fixed graph. We shall now give an example that shows that merely assuming optimality is regular and the uniform finite-memory determinacy of the future threshold games played on a given graph does not suffice to conclude the existence of Nash equilibria.

\begin{example}[\footnote{This example is based on an example communicated to the authors by Axel Haddad and Thomas Brihaye, which in turn is based on a construction in \cite{haddad}.}]
\label{example:mont}
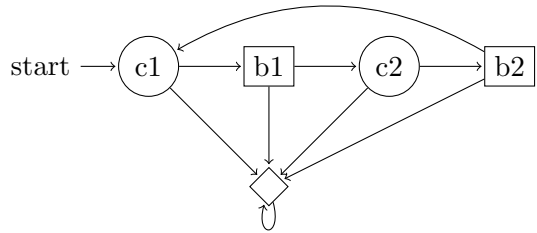
\begin{figure}[ht]
\centering
\begin{tikzpicture}[shorten >=1pt,node distance=1.6cm, auto]
  \node[draw, circle, initial] (x) {c1};
  \node[draw] (y)  [right of = x] {b1};
  \node[draw, circle] (x2)  [right of = y] {c2};
  \node[draw] (y2)  [right of = x2] {b2};
  \node[draw,diamond] (z) [below of = y] {};

\path[->] (x) edge node {} (y)
                (x) edge node {} (z)
                (y) edge node {} (x2)
                edge node {} (z)
                (x2) edge node {} (y2)
                edge node {} (z)
                (y2) edge [bend right] node {} (x)
                edge node {} (z)
                (z) edge [loop below] node {} ();
 \end{tikzpicture}
  \caption{The graph for the game in Example \ref{example:mont}}
 \label{fig:Mont-necessary}
 \end{figure}

The game $g$ in Figure~\ref{fig:Mont-necessary} involves Player $1$ ($2$) who owns the circle (box) vertices. Who owns the diamond is irrelevant. The payoff for Player $1$ ($2$) is the number of visits to a box (circle) vertex, if this number is finite, and is $-1$ otherwise.

{\bf Claim:} All future threshold games in $g$ are determined via positional strategies.
\begin{proof}
Let $s_1$ be the positional strategy where Player $1$ chooses $b1$ when in $c1$ and the diamond when in $c2$, let $s_2$ be the positional strategy where Player $1$ chooses the diamond in $c1$ and $b_2$ in $c_2$. Let $s_\infty$ be the positional strategy where Player $1$ always chooses the diamond. Likewise, let $t_1$ be the positional strategy of Player $2$ going to $c_2$ in $b_1$ and to diamond in $b_2$, let $t_2$ go to diamond in $b_1$ and to $c_1$ in $b_2$, and let $t_\infty$ always go to the diamond.

Consider some history $h$ ending in $c1$, which has seen $n$ occurrences of box vertices. By playing $s_1$, Player $1$ can win the future threshold game starting after $h$ for all threshold $k \leq n + 1$. His opponent can win for all higher thresholds by playing $t_2$. By symmetry of the game, we see that also for histories ending in $c2$, $b1$ or $b2$, the player can win all winnable future threshold games using one of $s_1$, $s_2$, $t_1$, $t_2$, and his opponent can win the remaining ones using the counterpart strategy.
\end{proof}

{\bf Claim:} The game $g$ has no Nash equilibrium (so in particular, neither optimal strategies nor finite memory Nash equilibrium).
\begin{proof}
In the run induced by a putative NE, one of the players has to choose the diamond at some point (to avoid payoff $-1$), but by postponing this choice to next time, the player can increase her payoff by $1$.
\end{proof}
\end{example}

\subsection{On uniform finite-memory determinacy}
We have seen that the existence of a uniform memory bound sufficient to win all winnable threshold games is necessary for the existence of finite-memory Nash equilibria. A typical example for a class of games failing this condition is found in mean-payoff parity games (which, being prefix-independent, have the optimality-is-regular property). The same example, however, also works as a discounted-payoff parity game.

\begin{example}
\label{ex:finite-unbounded-memory}
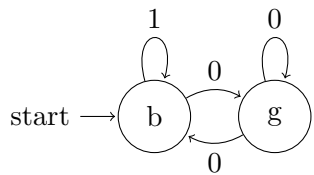
\begin{figure}[hb]
\centering
\begin{tikzpicture}[shorten >=1pt,node distance=1.6cm, auto]
  \node[state, initial] (x) {b};
  \node[state] (y)  [right of = x] {g};

\path[->] (x) edge [loop above] node {1} ()
                edge [bend left] node {0} (y)
                (y) edge [loop above] node {0} ()
                edge [bend left] node {0} (x);
 \end{tikzpicture}
  \caption{The graph for Example \ref{ex:finite-unbounded-memory}}
\label{fig:finite-unbounded-memory}
\end{figure}

Let $g$ be the one-player game in Figure~\ref{fig:finite-unbounded-memory}. The payoff of a run that visits the vertex $g$ infinitely often is the limit (inferior or superior) of the average payoff. It is zero if $g$ is visited finitely many times only. For any threshold $t \in \mathbb{R}$, if $t < 1$, the player has a winning finite-memory strategy: cycle $p$ times in $b$, where $p > \frac{1}{1-t}$, visit $g$ once, cycle $p$ times in $b$, and so on. If $t \geq 1$, the player has no winning strategy at all. So the thresholds games of $g$, and likewise for the future game of $g$, are finite-memory determined. The game has no finite-memory Nash equilibrium nonetheless, since the player can get a payoff as closed to $1$ as she wants, but not $1$.
\end{example}

Uniform finite-memory determinacy can sometimes be recovered by considering $\varepsilon$-versions instead: We partition the payoffs in blocks of size $\varepsilon$, and let the player be indifferent within the same block. Clearly any Nash equilibrium from the $\varepsilon$-discretized version yields an $\varepsilon$-Nash equilibrium of the original game. If the original preferences were prefix-independent, the modified preferences still are. Moreover, as there are now only finitely many relevant threshold games per graph, their uniform finite-memory determinacy follows from mere finite-memory determinacy. In such a situation, our result allows us to conclude that multi-player multi-outcome games have finite memory $\varepsilon$-Nash equilibria. For example, we obtain:

\begin{corollary}
Multi-player multi-outcome mean-payoff parity games have finite memory $\varepsilon$-Nash equilibria.
\end{corollary}

Here a multi-player multi-outcome mean-payoff parity game is understood to be a game where each player has payoff labels associated with it, and each vertex some priority. Players have some strict weak order preference on the pairs of the $\limsup$ or $\liminf$ of their average payoff on a prefix and the least priority seen infinitely often, which is consistent with the usual order on the payoff component (i.e.~getting more payoff while keeping the same priority is always better). A $\varepsilon$-Nash equilibrium is one where no player can improve by changing the priority, and no player can improve their payoff by more than $\varepsilon$.

\section{Applications}
\label{sec:applications}
We shall briefly mention two classes of games covered by our main theorem, but not by the results from \cite{depril,depril2}, (bounded) energy parity games and games with the lexicographic product of mean-payoff and reachability preferences. We leave the investigation whether winning conditions defined via $\mathrm{LTL}[\mathcal{F}]$ or $\mathrm{LTL}[\mathcal{D}]$ formulae \cite{kupferman2} match the criteria of Theorem \ref{thm:fmNE-many-games} to future work. Another area of prospective examples to explore are multi-dimensional objectives as studied e.g.~in \cite{raskin,raskin4}.

\subsection{Energy parity games}
\label{subsec:energyparity}
Energy games were first introduced in \cite{chakrabarti}: Two players take turns moving a token through a graph, while keeping track of the \emph{current energy level}, which will be some integer. Each move either adds or subtracts to the energy level, and if the energy level ever reaches $0$, the protagonist loses. These conditions were later combined with parity winning conditions in \cite{chatterjee4} to yield energy parity games as a model for a system specification that keeps track of gaining and spending of some resource, while simultaneously conforming to a parity specification.

In both \cite{chakrabarti} and \cite{chatterjee4} the energy levels are a priori not bounded from above. This is a problem for the applicability of Theorem \ref{thm:fmNE-one-game-bit}, since unbounded energy preferences do not have the optimality-is-regular property, as shown in Example~\ref{example:energynotoir} below.

\begin{example}
\label{example:energynotoir}
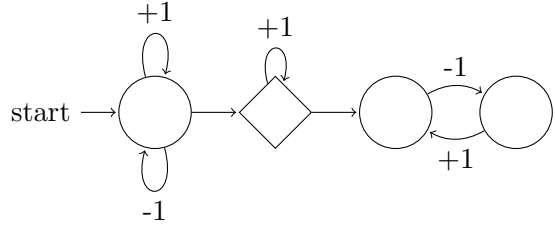
\begin{figure}[htb]
\centering
\begin{tikzpicture}[shorten >=1pt,node distance=1.6cm, auto]
  \node[state, initial, circle] (x) {};
  \node[state, diamond] (y)  [right of = x] {};
  \node[state, circle] (z1) [right of = y] {};
  \node[state, circle] (z2) [right of = z1] {};

\path[->] (x) edge [loop above] node {+1} ()
           (x) edge [loop below] node {-1} ()
           (x) edge  node {} (y)
           (y) edge node {} (z1)
           (y) edge [loop above] node {+1} ()
           (z1) edge [bend left] node {-1} (z2)
           (z2) edge [bend left] node {+1} (z1);
 \end{tikzpicture}
  \caption{The graph for Example \ref{example:energynotoir}}
\label{fig:energynotoir}
\end{figure}
Consider the game depicted in Figure \ref{fig:energynotoir}. The protagonist controls the diamond vertex, the opponent the circle vertices. Energy deltas are denoted on the edges.

{\bf Claim:} There is no finite automaton that decides whether the strategy of the protagonist that goes right straight-away on reaching diamond is optimal after some history in the energy game.
\begin{proof}
This strategy is optimal provided that the current energy level is not equal to the least energy level ever reached. In that case, taking the self-loop at the diamond vertex once and then going right would be preferable. But deciding whether the current energy level is equal to the least energy level essentially requires counting, and can thus not be done by a finite automaton.
\end{proof}
\end{example}

In \cite{bouyer}, two versions of bounded energy conditions were investigated: Either any energy gained in excess of the upper bound is just lost (as in e.g.~recharging a battery), or gaining energy in excess of the bound leads to a loss of the protagonist (as in e.g.~refilling a fuel tank without automatic spill-over prevention). We are only concerned with the former, and define our multi-player multi-outcome version as follows:

\begin{definition}
A multi-player multi-outcome energy parity game (MMEP game) is a game where each color is a tuple of ordered pairs in $\mathbb{Z} \times \mathbb{N}$, one pair for each player. The first (second) component of a pair is called an energy delta (a priority), noted $\delta_v^a$ ($\pi_v^a$).

The preferences are described as follows. Each player $a$ has an upper energy bound $E_{\max}^a \in \mathbb{N}$. The cumulative energy values $E_n^a$ for player $a$ in a run $\rho = v_0v_1\ldots$ are defined by $E_0^a := \min(E_{\max}^a,\delta_{v_0}^a)$ and $E_{n+1}^a := \min \{E_{\max}^a, E^a_n + \delta^a_{v_{n+1}}\}$. Player $a$ only cares about the least priority occurring infinitely many times together with $E^a = \min_{n \in \mathbb{N}} E^a_n$. He has some strict weak order $\prec_a$ on such pairs, which must respect $E < E' \Rightarrow (\pi, E') \nprec_a (\pi,E)$.
\end{definition}

The threshold games arising from MMEP games are the disjunctive bounded-energy parity games, which we define next:

\begin{definition}
A disjunctive bounded-energy parity game is a two-player win/lose game where colors are ordered pairs in $\mathbb{Z} \times \mathbb{N}$.

The winning condition is defined as follows. Let $E_{\max} \in \mathbb{N}$. The cumulative energy values $E_n$ in a run $\rho = v_0v_1\ldots$ are defined by $E_0 := \min(E_{\max},\delta_{v_0})$ and $E_{n+1} := \min \{E_{\max}, E_n + \delta_{v_{n+1}}\}$.

Given a run $\rho = v_0v_1\ldots$, we consider two values: The least priority $\pi$ occurring infinitely many times, and the least cumulative energy value reached, $E = \min_{n \in \mathbb{N}} E_n$. The winning condition is given by some family $(E_{\min}^i, P_i)_{i \in I}$ of energy thresholds and sets of priorities. Player $1$ wins iff there is some $i \in I$ with $E > E_{\min}^i$ and $\pi \in P_i$. We will write $B_{\min} = \min_{i \in I} E_{\min}^i$.
\end{definition}

The usual bounded-energy parity games are the special case where $|I| = 1$ (one can then of course easily rearrange the priorities such that the ones in $P_i$ are even, and the others odd). Some remarks on the relevance of disjunctions of winning conditions follow below on Page \pageref{subsub:disjunctive}.

\begin{theorem}
\label{theo:g2pwlep}
Disjunctive bounded-energy parity games are finite-memory determined, and $\log |E_{\max} - B_{\min}|$ bits of memory suffices for Player $1$, and $2\log |E_{\max} - B_{\min}|$ bits for Player $2$.
\begin{proof}
We take the product of $V$ with the set $Q := \{B_{\min}, \ldots, E_{\max}\} \times \{B_{\min}, \ldots, E_{\max}\}$, where there is an edge from $(v,e_0,e_1)$ to $(u,e_0',e_1')$ iff there an edge from $v$ to $u$ in the original graph, $e_0' := \min \{E_{\max}, \max \{B_{\min}, e_0 + \delta_{v}\}\}$ and $e_1' = \min \{e_1,e_0'\}$. Essentially, we keep track of both the current energy level and of the least energy level ever encountered as part of the vertices. Note that there never can be an edge from some $(v,e_0,e_1)$ to $(u,e_0',e_1')$ where $e_1' > e_1$.

Let $T_k := \{\pi \mid \exists i \ \pi \in P_i \wedge E^i_{\min} \leq B_{\min} + k\}$. Now we first consider the subgraph induced by the vertices of the form $(v,e_0,B_{\min})$; and then the parity game played on this subgraph with winning priorities $T_0$. As parity games admit positional strategies that win from every possible vertex, we can fix such strategies for both players on the subgraph. Let $A_0^0$ be the set of vertices in this subgraph where Player $1$ wins, and let $B_0^0$ be the set of vertices in this subgraph where Player $2$ wins.

We proceed by a reachability analysis: Let $A_0^{i+1}$ be $A_0^{i}$ together with all vertices controlled by Player $1$ that have an outgoing edge into $A_0^i$, and all vertices controlled by Player $2$ where all outgoing edges go into $A_0^i$. We extend the positional strategies of the players to $A_0^{i+1}$ by letting Player $1$ pick some witnessing outgoing edge, and Player $2$ some arbitrary edge. Likewise, we define $B_0^{i+1}$ be $B_0^i$ together with all vertices controlled by Player $1$ where all going edges go into $B_0^i$, and all vertices controlled by Player $2$ with some outgoing edge into $B_0^i$, and extend the strategies analogously.

It remains to define the strategies in the subgraph induced by $V \setminus (A_0^{|V|} \cup B_0^{|V|})$. In the next stage, we consider the subgraph of this subgraph induced by the vertices of the form $(v,e_0,B_{\min} + 1)$, and again consider a parity game played there, this time with winning priorities in $T_1$, and so on.

Iterating the parity-game and reachability analysis steps will yield positional optimal strategies for both players on the whole expanded graph.

Now consider the winning sets and strategies of both players: If Player $1$ wins from some vertex $(v,e_0,e_1)$, then he also wins from any $(v,e_0,e'_1)$ where $e_1 < e'_1$ -- for the only difference is the lowest energy level ever reached, which can only benefit, but not harm, Player $1$. Moreover, as $(v,e_0,e'_1)$ cannot be reached from $(v,e_0,e_1)$ at all, Player $1$ can safely play the same vertex $u$ in the original graph at $(v,e_0,e'_1)$ as he plays at $(v,e_0,e_1)$. For fixed $v$, $e_0$, let $e_1$ be minimal such that Player $1$ wins from $(v,e_0,e_1)$. Then we can change his strategy such that he plays the same vertex in the underlying graph from any $(v,e_0,e'_1)$.(\footnote{This trick does not work for Player 2, because
we would need to consider the maximal $e_1$ where she wins (instead of the
minimal one for Player 1), but then the backward induction from the
middle of the proof goes in the "wrong direction".})

By using $\log |E_{\max} - B_{\min}|$ bits of memory, Player $1$ can play his positional strategy from above in the original game. Likewise, Player $2$ can play her positional strategy from the expanded graph in the original game using $2 \log |E_{\max} - B_{\min}|$ bits of memory.
\end{proof}
\end{theorem}

We need one last simple lemma, and then will be able to apply Theorem \ref{thm:fmNE-one-game-bit} to energy parity games.

\begin{lemma}
\label{lem:epautomaticpiecewise}
The valuation-preference combinations in MMEP games are automatic-piecewise prefix-independent with at most $nE^2$ pieces, where $n$ is the size of the graph and $E$ is the maximum difference between the energy maximum and the energy minimum for some player.
\begin{proof}
The pieces are defined by the current vertex, the current energy level, and the least ever energy level, i.e.~the values $E_n^a$ and $\min_{j \leq n} E_n^a$. As energy is bounded, both enrgy levels can easily be computed by a finite automaton. If $h$ and $h'$  end with the same vertex and share the same current energy level and least energy level, then the least energy level reached in $h p$ is equal to the least energy level reached in $h' p$. As the least priority seen infinitely many times depends only on the tail, but never on the finite prefix, we see that for $ h$ and $ h'$ being the same piece, $h p$ and $h' p$ are interchangeable for the player.
\end{proof}
\end{lemma}

\begin{corollary}
All multiplayer multioutcome energy parity games have Nash equilibria in finite memory strategies. Let $A$ be the set of players, $n$ the size of the graph and let $E$ be the maximum difference between the energy maximum and the energy minimum for some player. Then $2|A|E^2n\log E^2n + 1$ bits of memory suffice.
\begin{proof}
By Theorem \ref{theo:g2pwlep} and Lemma \ref{lem:epautomaticpiecewise} the prerequisites of Theorem \ref{thm:fmNE-one-game-bit} are given. The Mont condition is trivially true, as there are no infinite ascending chains in the preferences. From Theorem \ref{theo:g2pwlep} we see that the parameter $T$ in Theorem \ref{thm:fmNE-one-game-bit} can be chosen as $2\log E$. By Lemma \ref{lem:epautomaticpiecewise} we can chose $D = \log E^2n$ and $k = E^2n$. First note that the claim holds for $n = 1$ or $E = 1$ (by positional determinacy of parity games). Second, for $2 \leq n$ and $2 \leq E$ we have $\log nE^2 \leq nE^2 \log n$, so straightforward calculus shows the general claim.
\end{proof}
\end{corollary}

\subsubsection*{Algorithmic considerations}
The proof of Theorem \ref{theo:g2pwlep} immediately gives rise to an algorithm computing the winning strategies in disjunctive bounded-energy parity game while using an oracle for winning strategies in parity games. Using e.g.~the algorithm for solving parity games from \cite{paterson}, which has a runtime of $n^{O(\sqrt{n})}$, we obtain a runtime of $(nE)^{O(\sqrt{nE})}$, if we set $E := |E_{\max} - B_{\min}|$. Unfortunately, only the binary representation of $E$ will need to be present in the input -- $E$ itself can easily be exponential in the size of the input.

If we assume $W$ to be fixed\footnote{Which is poly-time equivalent to $W$ being given in unary.}, we arrive at a Cook reduction of solving disjunctive bounded-energy parity games to solving parity games. This in particular implies that the decision problem for disjunctive bounded-energy parity games with bounded weights is in $\textrm{P}^{(\textrm{UP} \cap \textrm{co-UP})}$.

\subsubsection*{Disjunctions of winning conditions}
\label{subsub:disjunctive}
Parity and Muller conditions are easily seen to be closed under conjunction and disjunction, as these just correspond to intersection and union respectively of the relevant sets of winning priorities, or sets of winning sets of vertices visited infinitely many often. As long as just a single notion of energy (or respectively payoff) is available, likewise energy, mean-payoff and discounted payoff conditions are closed under conjunction and disjunction, as here these logical connectives just correspond to minimum and maximum on the threshold values. Subsequently, despite the high relevance of boolean operations on winning conditions, it is unsurprising that they have received little attention in the literature so far.

For energy parity conditions, which are themselves a conjunction of parity and energy conditions, the considerations above immediately imply closure under conjunction. The disjunction of two energy parity conditions, however, is not necessarily equivalent to an energy parity condition. In fact, we even see a qualitative difference with respect to the memory requirements for Player $2$: It was shown by \name{Chatterjee} and \name{Doyen} that in an (unbounded) energy parity game, if Player $2$ has a winning strategy, she has a positional one. This translates directly to the corresponding result for bounded energy parity games. For disjunctive energy parity games, Player $2$ might require memory to win though:

\begin{example}
\label{example:memorydisjunctive}
In the energy parity game depicted in Figure \ref{fig:memorydisjunctive}, Player $2$ controls all vertices. Player $1$ wins with any priority, if the energy stays above $5$, and wins with priority $0$ if the energy stays above $0$. Each vertex is marked with $\pi / \delta$, where $\pi$ is the priority and $\delta$ the energy delta.

Any positional strategy of Player $2$ is winning for Player $1$, but by e.g.~alternating the sole choice she has, Player $2$ can win using one bit of memory.

\begin{figure}
\centering
\begin{tikzpicture}[shorten >=1pt,node distance=1.6cm, auto]
  \node[draw, circle, initial] (x) {$1 / +5$};
  \node[draw, circle] (y)  [right of = x] {$1 / 0$};
  \node[draw, circle] (y2)  [right of = y] {$1 / +5$};
  \node[draw,circle] (y3)  [below of = y2] {$1 / -5$};
  \node[draw,circle] (z) [below of = y] {$0 / 0$};

\path[->] (x) edge node {} (y)
                (y) edge node {} (y2)
                (y) edge node {} (z)
                (z) edge node {} (y)
                (y2) edge node {} (y3)
                (y3) edge node {} (y)
;
 \end{tikzpicture}
  \caption{The graph for the game in Example \ref{example:memorydisjunctive}}
 \label{fig:memorydisjunctive}
 \end{figure}
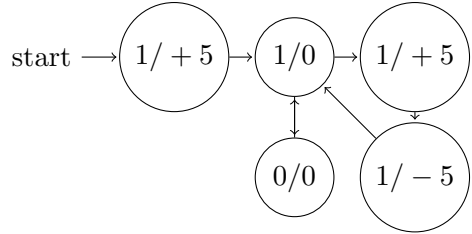
\end{example}

\subsection{Lexicographic product of Mean-payoff and reachability}
\label{subsec:lexreachability}
In our second example, each player has both a mean-payoff goal and a reachability objective. Maximizing the mean-payoff, however, takes precedence, and the reachability objective only becomes relevant as a tie-breaker; i.e.~we consider the lexicographic combination of the mean-payoff preferences with the reachability objective. These preferences are of particular interest as they cannot be expressed via a payoff function\footnote{The von Neumann Morgenstern utility theorem \cite{neumann} does not apply, as the continuity axiom is not satisfied.}, hence are an example for why considering preferences instead of payoff functions is useful.

\begin{proposition}
The lexicographic product of mean-payoff preferences and a reachability objective cannot be expressed as a payoff function.
\begin{proof}
It is straight-forward to construct an example of a game where any mean payoff in $\uint$ in any combination with reaching or not reaching the reachability objective is realizable. If there were an equivalent payoff function for this game, it would induce an order embedding $\iota$ of $\uint \times_{\textrm{lex}} \{0,1\}$ into $\mathbb{R}$. As $(x, 0) \prec (x,1)$ for any $x \in \uint$, we would find that $\iota(x,0) < \iota(x,1)$. Thus, there has to be some rational $q_x$ with $\iota(x,0) < q_x < \iota(x,1)$. Moreover, as $(x,1) \prec (y,0)$ for $x < y$, we find that $q_x \neq q_y$ for $x \neq y$. But then $x \mapsto q_x$ would be an injection from $\uint$ to $\mathbb{Q}$, which cannot exist for reasons of cardinality.
\end{proof}
\end{proposition}

To deal with lexicographic products, the following will be very useful:
 \begin{lemma}
The weak OIR property is preserved by lexicographic products.
\begin{proof}
Let $\prec_1$ and $\prec_2$ have the OIR property, and let $p,q \in C^\omega \cap \textrm{reg}$. By assumption there exist regular sets $M_1$, $M_1'$, and $M_2$ such that
for all $h \in \mathcal{H}$ we have $(h \in M_1 \Leftrightarrow h\hat{\,}p \prec_1 h\hat{\,}q )$ and $(h \in M_1' \Leftrightarrow h\hat{\,}q \prec_1 h\hat{\,}p )$ and $(h \in M_2 \Leftrightarrow h\hat{\,}p \prec_2 h\hat{\,}q)$. So $h\hat{\,}p (\prec_1\times_{lex}\prec_2) h\hat{\,}q$ iff $h \in M_1\cup (M_2 \setminus M_1')$.
\end{proof}
\end{lemma}

By Proposition \ref{prop:weak-oir-oir}, it only remains to show that the induced threshold games are uniformly finite-memory determined. We show a slightly stronger result:

\begin{lemma}
In the threshold games where the preferences are the lexicographic product of mean-payoff and reachability, either Player $1$ has a winning strategy using one bit of memory, or Player $2$ has a positional winning strategy.
\begin{proof}
If the threshold is of the form $(x, 1)$, then the reachability component is irrelevant, and the game is equivalent to the threshold game of a mean-payoff game. As these are positionally determined, the claim is immediate.

If the threshold is of the form $(x,0)$, then Player $1$ has two ways of winning: Either get mean-payoff strictly more than $x$, or reach the target set and obtain a mean-payoff of at least $x$. We can consider for each vertex the value of the mean-payoff game starting there. Player $1$ has a positional strategy that obtains more than $x$ mean-payoff from all vertices where this is possible, and this strategy ensures that this region is never left. If this region includes the starting vertex, we are done. Otherwise, note that Player $1$ cannot enter this region from the outside, and Player $2$ has no incentive to ever enter this region from the outside. Thus, we can restrict our attention to the game played on the induced subgraph on the complement.

In the remaining game, we consider those vertices in the target set where the mean-payoff obtainable is $x$. From all vertices where Player $1$ can force the play to reach one of these, he can win with a strategy using a single bit of memory: Play towards such a vertex in a positional way, then flip the bit and follow a positional strategy ensuring mean-payoff at least $x$. Again, if the starting vertex is covered, we are done. Else, note that Player $1$ cannot reach the region from the outside, and Player $2$ has no incentive to.

In the game remaining after the second step, we again compute the obtainable mean-payoff values. Player $2$'s refusal to enter the region in round $2$ might increase the mean-payoff Player $1$ can obtain above $x$, and thus let him win after all. In any case, if both players follow positional optimal strategies for mean-payoff games in the remaining part, they will win the mean-payoff plus reachability game if they can at all.
\end{proof}
\end{lemma}

\begin{corollary}
The multi-player multi-outcome games where preferences are lexicographic products of mean-payoff and reachability objectives have finite-memory Nash equilibria.
\end{corollary}

\section*{Acknowledgements}

We are indebted to  Nathana{\"e}l Fijalkow for presenting our work on our behalf at the Strategic Reasoning Workshop, as well as to Axel Haddad and Thomas Brihaye for a very useful conversation.

\bibliographystyle{eptcs}
\bibliography{../../../../spieltheorie}

\begin{thebibliography}{10}
\providecommand{\bibitemdeclare}[2]{}
\providecommand{\surnamestart}{}
\providecommand{\surnameend}{}
\providecommand{\urlprefix}{Available at }
\providecommand{\url}[1]{\texttt{#1}}
\providecommand{\href}[2]{\texttt{#2}}
\providecommand{\urlalt}[2]{\href{#1}{#2}}
\providecommand{\doi}[1]{doi:\urlalt{http://dx.doi.org/#1}{#1}}
\providecommand{\bibinfo}[2]{#2}

\bibitemdeclare{article}{kupferman2}
\bibitem{kupferman2}
\bibinfo{author}{S.~\surnamestart Almagor\surnameend},
  \bibinfo{author}{U.~\surnamestart Boker\surnameend} \&
  \bibinfo{author}{O~\surnamestart Kupferman\surnameend}
  (\bibinfo{year}{2016}): \emph{\bibinfo{title}{Formalizing and reasoning about
  quality}}.
\newblock {\sl \bibinfo{journal}{Journal of the ACM}}, \doi{10.1145/2875421}.

\bibitemdeclare{incollection}{bouyer}
\bibitem{bouyer}
\bibinfo{author}{Patricia \surnamestart Bouyer\surnameend},
  \bibinfo{author}{Uli \surnamestart Fahrenberg\surnameend},
  \bibinfo{author}{Kim~G. \surnamestart Larsen\surnameend},
  \bibinfo{author}{Nicolas \surnamestart Markey\surnameend} \&
  \bibinfo{author}{Ji\v{r}\'i \surnamestart Srba\surnameend}
  (\bibinfo{year}{2008}): \emph{\bibinfo{title}{Infinite Runs in Weighted Timed
  Automata with Energy Constraints}}.
\newblock In \bibinfo{editor}{Franck \surnamestart Cassez\surnameend} \&
  \bibinfo{editor}{Claude \surnamestart Jard\surnameend}, editors: {\sl
  \bibinfo{booktitle}{Formal Modeling and Analysis of Timed Systems}}, {\sl
  \bibinfo{series}{Lecture Notes in Computer Science}} \bibinfo{volume}{5215},
  \bibinfo{publisher}{Springer Berlin Heidelberg}, pp. \bibinfo{pages}{33--47},
  \doi{10.1007/978-3-540-85778-5\_4}.

\bibitemdeclare{inproceedings}{depril3}
\bibitem{depril3}
\bibinfo{author}{Thomas \surnamestart Brihaye\surnameend},
  \bibinfo{author}{Veroniqu\'e \surnamestart Bruy\`ere\surnameend} \&
  \bibinfo{author}{Julie \surnamestart {De Pril}\surnameend}
  (\bibinfo{year}{2010}): \emph{\bibinfo{title}{Equilibria in quantitative
  reachability games}}.
\newblock In: {\sl \bibinfo{booktitle}{Proc. of CSR}}, {\sl
  \bibinfo{series}{LNCS}} \bibinfo{volume}{6072},
  \bibinfo{publisher}{Springer}, \doi{10.1007/978-3-642-13182-0\_7}.

\bibitemdeclare{inproceedings}{depril2}
\bibitem{depril2}
\bibinfo{author}{Thomas \surnamestart Brihaye\surnameend},
  \bibinfo{author}{Julie \surnamestart {De Pril}\surnameend} \&
  \bibinfo{author}{Sven \surnamestart Schewe\surnameend}
  (\bibinfo{year}{2013}): \emph{\bibinfo{title}{Multiplayer Cost Games with
  Simple {N}ash Equilibria}}.
\newblock In: {\sl \bibinfo{booktitle}{Logical Foundations of Computer
  Science}}, \bibinfo{series}{LNCS}, pp. \bibinfo{pages}{59--73},
  \doi{10.1007/978-3-642-35722-0\_5}.

\bibitemdeclare{misc}{haddad}
\bibitem{haddad}
\bibinfo{author}{Thomas \surnamestart Brihaye\surnameend},
  \bibinfo{author}{Gilles \surnamestart Geeraerts\surnameend},
  \bibinfo{author}{Axel \surnamestart Haddad\surnameend} \&
  \bibinfo{author}{Benjamin \surnamestart Monmege\surnameend}
  (\bibinfo{year}{2014}): \emph{\bibinfo{title}{To Reach or not to Reach?
  {E}fficient Algorithms for Total-Payoff Games}}.
\newblock \bibinfo{howpublished}{arXiv 1407.5030}.
\newblock \urlprefix\url{http://arxiv.org/abs/1407.5030}.

\bibitemdeclare{inproceedings}{bulling}
\bibitem{bulling}
\bibinfo{author}{Nils \surnamestart Bulling\surnameend} \&
  \bibinfo{author}{Valentin \surnamestart Goranko\surnameend}
  (\bibinfo{year}{2013}): \emph{\bibinfo{title}{How to be both rich and happy:
  Combining quantitative and qualitative strategic reasoning about multi-player
  games (Extended Abstract)}}.
\newblock In: {\sl \bibinfo{booktitle}{Proc. of Strategic Reasoning}},
  \doi{10.4204/EPTCS.112.8}.
\newblock \urlprefix\url{http://www.arxiv.org/abs/1303.0789}.

\bibitemdeclare{incollection}{chakrabarti}
\bibitem{chakrabarti}
\bibinfo{author}{Arindam \surnamestart Chakrabarti\surnameend},
  \bibinfo{author}{Luca \surnamestart de~Alfaro\surnameend},
  \bibinfo{author}{Thomas~A. \surnamestart Henzinger\surnameend} \&
  \bibinfo{author}{Mari\"elle \surnamestart Stoelinga\surnameend}
  (\bibinfo{year}{2003}): \emph{\bibinfo{title}{Resource Interfaces}}.
\newblock In \bibinfo{editor}{Rajeev \surnamestart Alur\surnameend} \&
  \bibinfo{editor}{Insup \surnamestart Lee\surnameend}, editors: {\sl
  \bibinfo{booktitle}{Embedded Software}}, {\sl \bibinfo{series}{Lecture Notes
  in Computer Science}} \bibinfo{volume}{2855}, \bibinfo{publisher}{Springer
  Berlin Heidelberg}, pp. \bibinfo{pages}{117--133},
  \doi{10.1007/978-3-540-45212-6\_9}.

\bibitemdeclare{article}{chatterjee4}
\bibitem{chatterjee4}
\bibinfo{author}{Krishnendu \surnamestart Chatterjee\surnameend} \&
  \bibinfo{author}{Laurent \surnamestart Doyen\surnameend}
  (\bibinfo{year}{2012}): \emph{\bibinfo{title}{Energy parity games}}.
\newblock {\sl \bibinfo{journal}{Theor. Comput. Sci.}} \bibinfo{volume}{458},
  pp. \bibinfo{pages}{49--60}, \doi{10.1016/j.tcs.2012.07.038}.

\bibitemdeclare{inproceedings}{raskin4}
\bibitem{raskin4}
\bibinfo{author}{Lorenzo \surnamestart Clemente\surnameend} \&
  \bibinfo{author}{Jean{-}Fran{\c{c}}ois \surnamestart Raskin\surnameend}
  (\bibinfo{year}{2015}): \emph{\bibinfo{title}{Multidimensional beyond
  Worst-Case and Almost-Sure Problems for Mean-Payoff Objectives}}.
\newblock In: {\sl \bibinfo{booktitle}{30th Annual {ACM/IEEE} Symposium on
  Logic in Computer Science, {LICS} 2015, Kyoto, Japan, July 6-10, 2015}}, pp.
  \bibinfo{pages}{257--268}, \doi{10.1109/LICS.2015.33}.

\bibitemdeclare{article}{fijalkow2}
\bibitem{fijalkow2}
\bibinfo{author}{Nathana\"el \surnamestart Fijalkow\surnameend} \&
  \bibinfo{author}{Martin \surnamestart Zimmermann\surnameend}
  (\bibinfo{year}{2014}): \emph{\bibinfo{title}{Parity and Streett Games with
  Costs}}.
\newblock {\sl \bibinfo{journal}{Logical Methods in Computer Science}}
  \bibinfo{volume}{10}(\bibinfo{number}{2}), \doi{10.2168/LMCS-10(2:14)2014}.

\bibitemdeclare{inproceedings}{gimbert}
\bibitem{gimbert}
\bibinfo{author}{Hugo \surnamestart Gimbert\surnameend} \&
  \bibinfo{author}{Wies{\l}aw \surnamestart Zielonka\surnameend}
  (\bibinfo{year}{2005}): \emph{\bibinfo{title}{Games Where You Can Play
  Optimally Without Any Memory}}.
\newblock In \bibinfo{editor}{Mart{\'i}n \surnamestart Abadi\surnameend} \&
  \bibinfo{editor}{Luca \surnamestart de~Alfaro\surnameend}, editors: {\sl
  \bibinfo{booktitle}{CONCUR 2005. Proceedings}}, \bibinfo{publisher}{Springer
  Berlin Heidelberg}, \bibinfo{address}{Berlin, Heidelberg}, pp.
  \bibinfo{pages}{428--442}, \doi{10.1007/11539452\_33}.
\newblock \urlprefix\url{http://dx.doi.org/10.1007/11539452\_33}.

\bibitemdeclare{incollection}{ummels2}
\bibitem{ummels2}
\bibinfo{author}{Erich \surnamestart Gr\"adel\surnameend} \&
  \bibinfo{author}{M.~\surnamestart Ummels\surnameend} (\bibinfo{year}{2008}):
  \emph{\bibinfo{title}{Solution concepts and algorithms for Infinite
  multiplayer game}}.
\newblock In \bibinfo{editor}{K.~\surnamestart Apt\surnameend} \&
  \bibinfo{editor}{R.~\surnamestart van Rooij\surnameend}, editors: {\sl
  \bibinfo{booktitle}{New Perspectives on Games and Interaction}}, {\sl
  \bibinfo{series}{Texts in Logic and Games}}~\bibinfo{volume}{4},
  \bibinfo{publisher}{Amsterdam University Press}, pp.
  \bibinfo{pages}{151--178}.

\bibitemdeclare{inproceedings}{gurevich2}
\bibitem{gurevich2}
\bibinfo{author}{Yuri \surnamestart Gurevich\surnameend} \&
  \bibinfo{author}{L.~\surnamestart Harrington\surnameend}
  (\bibinfo{year}{1982}): \emph{\bibinfo{title}{Trees, automata and games}}.
\newblock In: {\sl \bibinfo{booktitle}{Proc. STOC}},
  \doi{10.1145/800070.802177}.

\bibitemdeclare{article}{paterson}
\bibitem{paterson}
\bibinfo{author}{Marcin \surnamestart Jurdzinski\surnameend},
  \bibinfo{author}{Mike \surnamestart Paterson\surnameend} \&
  \bibinfo{author}{Uri \surnamestart Zwick\surnameend} (\bibinfo{year}{2008}):
  \emph{\bibinfo{title}{A Deterministic Subexponential Algorithm for Solving
  Parity Games}}.
\newblock {\sl \bibinfo{journal}{SIAM J. Comput.}}
  \bibinfo{volume}{38}(\bibinfo{number}{4}), pp. \bibinfo{pages}{1519--1532},
  \doi{10.1137/070686652}.

\bibitemdeclare{inproceedings}{kupferman}
\bibitem{kupferman}
\bibinfo{author}{Orna \surnamestart Kupferman\surnameend}
  (\bibinfo{year}{2016}): \emph{\bibinfo{title}{On High-Quality Synthesis}}.
\newblock In \bibinfo{editor}{S.~Alexander \surnamestart Kulikov\surnameend} \&
  \bibinfo{editor}{J.~Gerhard \surnamestart Woeginger\surnameend}, editors:
  {\sl \bibinfo{booktitle}{11th International Computer Science Symposium in
  Russia, CSR 2016}}, \bibinfo{publisher}{Springer International Publishing},
  pp. \bibinfo{pages}{1--15}, \doi{10.1007/978-3-319-34171-2\_1}.

\bibitemdeclare{article}{leroux3}
\bibitem{leroux3}
\bibinfo{author}{St\'ephane \surnamestart Le~Roux\surnameend}
  (\bibinfo{year}{2013}): \emph{\bibinfo{title}{Infinite Sequential {N}ash
  Equilibria}}.
\newblock {\sl \bibinfo{journal}{Logical Methods in Computer Science}}
  \bibinfo{volume}{9}(\bibinfo{number}{2}), \doi{10.2168/LMCS-9(2:3)2013}.

\bibitemdeclare{inproceedings}{paulyleroux2}
\bibitem{paulyleroux2}
\bibinfo{author}{St{\'e}phane \surnamestart Le~Roux\surnameend} \&
  \bibinfo{author}{Arno \surnamestart Pauly\surnameend} (\bibinfo{year}{2014}):
  \emph{\bibinfo{title}{Infinite Sequential Games with Real-valued Payoffs}}.
\newblock In: {\sl \bibinfo{booktitle}{CSL-LICS '14}},
  \bibinfo{publisher}{ACM}, pp. \bibinfo{pages}{62:1--62:10},
  \doi{10.1145/2603088.2603120}.

\bibitemdeclare{incollection}{paulyleroux3-cie}
\bibitem{paulyleroux3-cie}
\bibinfo{author}{St\'ephane \surnamestart Le~Roux\surnameend} \&
  \bibinfo{author}{Arno \surnamestart Pauly\surnameend} (\bibinfo{year}{2015}):
  \emph{\bibinfo{title}{Weihrauch Degrees of Finding Equilibria in Sequential
  Games}}.
\newblock In \bibinfo{editor}{Arnold \surnamestart Beckmann\surnameend},
  \bibinfo{editor}{Victor \surnamestart Mitrana\surnameend} \&
  \bibinfo{editor}{Mariya \surnamestart Soskova\surnameend}, editors: {\sl
  \bibinfo{booktitle}{Evolving Computability}}, {\sl \bibinfo{series}{Lecture
  Notes in Computer Science}} \bibinfo{volume}{9136},
  \bibinfo{publisher}{Springer}, pp. \bibinfo{pages}{246--257},
  \doi{10.1007/978-3-319-20028-6\_25}.

\bibitemdeclare{inproceedings}{paulyleroux4-sr}
\bibitem{paulyleroux4-sr}
\bibinfo{author}{St\'ephane \surnamestart Le~Roux\surnameend} \&
  \bibinfo{author}{Arno \surnamestart Pauly\surnameend} (\bibinfo{year}{2016}):
  \emph{\bibinfo{title}{Extending Finite Memory Determinacy to Multiplayer
  Games}}.
\newblock In \bibinfo{editor}{Alessio \surnamestart Lomuscio\surnameend} \&
  \bibinfo{editor}{Moshe~Y. \surnamestart Vardi\surnameend}, editors: {\sl
  \bibinfo{booktitle}{{\rm Proc.} Strategic Reasoning, {\rm 2016}}}, {\sl
  \bibinfo{series}{EPTCS}} \bibinfo{volume}{218}, \bibinfo{publisher}{Open
  Publishing Association}, pp. \bibinfo{pages}{27--40},
  \doi{10.4204/EPTCS.218.3}.

\bibitemdeclare{inproceedings}{mertens}
\bibitem{mertens}
\bibinfo{author}{Jean~Fran\c{c}ois \surnamestart Mertens\surnameend}
  (\bibinfo{year}{1987}): \emph{\bibinfo{title}{Repeated Games}}.
\newblock In: {\sl \bibinfo{booktitle}{Proc. Internat. Congress
  Mathematicians}}, \bibinfo{publisher}{American Mathematical Society}, pp.
  \bibinfo{pages}{1528--1577}.

\bibitemdeclare{book}{neumann}
\bibitem{neumann}
\bibinfo{author}{John \surnamestart von Neumann\surnameend} \&
  \bibinfo{author}{Oskar \surnamestart Morgenstern\surnameend}
  (\bibinfo{year}{1944}): \emph{\bibinfo{title}{Theory of Games and Economic
  Behavior}}.
\newblock \bibinfo{publisher}{Princeton University Press}.

\bibitemdeclare{inproceedings}{soumya}
\bibitem{soumya}
\bibinfo{author}{Soumya \surnamestart Paul\surnameend} \&
  \bibinfo{author}{Sunil \surnamestart Simon\surnameend}
  (\bibinfo{year}{2009}): \emph{\bibinfo{title}{{Nash Equilibrium in
  Generalised Muller Games}}}.
\newblock In \bibinfo{editor}{Ravi \surnamestart Kannan\surnameend} \&
  \bibinfo{editor}{K.~Narayan \surnamestart Kumar\surnameend}, editors: {\sl
  \bibinfo{booktitle}{IARCS Annual Conference on Foundations of Software
  Technology and Theoretical Computer Science}}, {\sl \bibinfo{series}{Leibniz
  International Proceedings in Informatics (LIPIcs)}}~\bibinfo{volume}{4},
  \bibinfo{publisher}{Schloss Dagstuhl--Leibniz-Zentrum fuer Informatik},
  \bibinfo{address}{Dagstuhl, Germany}, pp. \bibinfo{pages}{335--346},
  \doi{10.4230/LIPIcs.FSTTCS.2009.2330}.
\newblock \urlprefix\url{http://drops.dagstuhl.de/opus/volltexte/2009/2330}.

\bibitemdeclare{phdthesis}{depril}
\bibitem{depril}
\bibinfo{author}{Julie~De \surnamestart Pril\surnameend}
  (\bibinfo{year}{2013}): \emph{\bibinfo{title}{Equilibria in Multiplayer Cost
  Games}}.
\newblock Ph.D. thesis, \bibinfo{school}{Universit\'e de Mons}.

\bibitemdeclare{article}{thuijsman}
\bibitem{thuijsman}
\bibinfo{author}{Frank \surnamestart Thuijsman\surnameend} \&
  \bibinfo{author}{Thirukkannamangai E.~S. \surnamestart Raghavan\surnameend}
  (\bibinfo{year}{1997}): \emph{\bibinfo{title}{Perfect information stochastic
  games and related classes}}.
\newblock {\sl \bibinfo{journal}{International Journal of Game Theory}}
  \bibinfo{volume}{26}(\bibinfo{number}{3}), pp. \bibinfo{pages}{403--408},
  \doi{10.1007/BF01263280}.
\newblock \urlprefix\url{http://dx.doi.org/10.1007/BF01263280}.

\bibitemdeclare{article}{raskin}
\bibitem{raskin}
\bibinfo{author}{Yaron \surnamestart Velner\surnameend},
  \bibinfo{author}{Krishnendu \surnamestart Chatterjee\surnameend},
  \bibinfo{author}{Laurent \surnamestart Doyen\surnameend},
  \bibinfo{author}{Thomas~A. \surnamestart Henzinger\surnameend},
  \bibinfo{author}{Alexander \surnamestart Rabinovich\surnameend} \&
  \bibinfo{author}{Jean-Fran\c{c}ois \surnamestart Raskin\surnameend}
  (\bibinfo{year}{2015}): \emph{\bibinfo{title}{The complexity of
  multi-mean-payoff and multi-energy games}}.
\newblock {\sl \bibinfo{journal}{Information and Computation}}
  \bibinfo{volume}{241}, pp. \bibinfo{pages}{177 -- 196},
  \doi{10.1016/j.ic.2015.03.001}.
\newblock
  \urlprefix\url{http://www.sciencedirect.com/science/article/pii/S0890540115000164}.

\end{thebibliography}

\end{document}